\documentclass[times,twocolumn]{aastex62}

\hypersetup{colorlinks=true,citecolor=black,linkcolor=black,urlcolor=black,filecolor=black,breaklinks=true}

\usepackage{amsmath}
\usepackage{txfonts,fleqn}

\usepackage{tikz}
\usetikzlibrary{calc}

\usepackage{acronym}
\newacro{BH}{black hole}
\newacroplural{BH}[BHs]{black holes}
\newcommand{\BH}{\ac{BH}}
\newcommand{\BHs}{\acp{BH}}
\newacro{DF}{distribution function}

\newcommand{\rd}{\mathrm{d}}
\newcommand{\re}{\mathrm{e}}
\newcommand{\ri}{\mathrm{i}}
\newcommand{\MBH}{M_{\bullet}}
\newcommand{\Msun}{M_{\odot}}
\newcommand{\mstar}{m_{\star}}
\newcommand{\emin}{e_{\mathrm{min}}}
\newcommand{\emax}{e_{\mathrm{max}}}
\newcommand{\Ekep}{E_{\mathrm{Kep}}}
\newcommand{\rh}{r_{\mathrm{h}}}
\newcommand{\pc}{\mathrm{pc}}
\newcommand{\amin}{a_{\mathrm{min}}}
\newcommand{\amax}{a_{\mathrm{max}}}
\newcommand{\rel}{\textsc{gr}}
\newcommand{\pw}{\star}
\newcommand{\yr}{\mathrm{yr}}
\newcommand{\ImPart}{\mathrm{Im}}
\newcommand{\RePart}{\mathrm{Re}}
\renewcommand*{\vec}[1]     {\boldsymbol{#1}}
\newcommand*{\uvec}[1]      {\hat{\vec{#1}}}
\newcommand*{\dvec}[1]      {\dot{\vec{#1}}}
\newcommand{\Mtot}{M_{\mathrm{tot}}}
\newcommand{\Mstar}{M_{\star}}
\newcommand{\PN}{\mathrm{PN}}
\newcommand{\Kep}{\mathrm{Kep}}

\newcommand{\all}{\mathrm{all}}
\newcommand{\self}{\mathrm{self}}

\newcommand{\ellmax}{\ell_{\mathrm{max}}}
\newcommand{\KG}{K_{\mathrm{G}}}

\newcommand{\tmax}{t_{\mathrm{max}}}

\newcommand{\Etot}{E_{\mathrm{tot}}}
\newcommand{\Etotdirect}{E_{\mathrm{tot}}^{\mathrm{direct}}}
\newcommand{\Ltot}{\vec{L}_{\mathrm{tot}}}
\newcommand{\taudyn}{\tau_{\mathrm{dyn}}}

\newcommand{\avforces}{\langle \#\text{ forces} \rangle}
\newcommand{\direct}{\mathrm{direct}}
\newcommand{\tvOmega}{\tilde{\vec{\Omega}}}
\newcommand{\dphiinv}{\mathrm{d} \phi^{-1}}
\newcommand{\Kmax}{K_{\mathrm{max}}}
\newcommand{\ad}{\mathrm{ad}}
\newcommand {\avint}[1] {    \oint\frac{\rd M_{#1}}{2\pi}}
\newcommand {\av}[2] {    \left\langle{#2}\right\rangle_{#1}}

\usepackage{soul} \usepackage{xcolor}

\shorttitle{Secular Dynamics and Multipole Expansion}
\shortauthors{Fouvry, Dehnen, Tremaine, Bar-Or}

\begin{document}

\title{Secular Dynamics around a Supermassive Black Hole via Multipole Expansion}

\author{Jean-Baptiste Fouvry}
\affiliation{Institut d'Astrophysique de Paris, UMR 7095, CNRS, Sorbonne Universit\'e, 98 bis Boulevard Arago, 75014 Paris, France}
\affiliation{Institute for Advanced Study, Princeton, NJ, 08540, USA}
\author{Walter Dehnen}
\affiliation{Astronomisches Rechen-Institut, Zentrum f{\"u}r Astronomie der Universit{\"a}t Heidelberg, M{\"o}nchhofstr 12-14, 69120 Heidelberg, Germany}
\affiliation{Universit{\"a}ts-Sternwarte der Ludwig-Maximilians-Universit{\"a}t, Scheinerstrasse 1, 81679 M{\"u}nchen, Germany}
\affiliation{School of Physics and Astronomy, University of Leicester, University Road, Leicester LE1 7RH, UK}
\author{Scott Tremaine}
\affiliation{Institute for Advanced Study, Princeton, NJ, 08540, USA}
\affiliation{Canadian Institute for Theoretical Astrophysics, University of Toronto, 60 St.\ George Street,
Toronto, ON M5S 3H8, Canada}
\author{Ben Bar-Or}
\affiliation{Institute for Advanced Study, Princeton, NJ, 08540, USA}


\begin{abstract}
In galactic nuclei, the gravitational potential is dominated by the central supermassive black hole, so stars follow quasi-Keplerian orbits. These orbits are distorted by gravitational forces from other stars, leading to long-term orbital relaxation. The direct numerical study of these processes is challenging because the fast orbital motion imposed by the central black hole requires very small timesteps. An alternative approach, pioneered by Gau\ss{}, is to use  the secular approximation of smearing out the $N$ stars over their Keplerian orbits, using $K$ nodes along each orbit.
In this study we propose three novel improvements to this method. First, we re-formulate the discretisation of the rates of change of the variables describing the orbital states to ensure that all conservation laws are exactly satisfied. Second, we replace the pairwise sum over nodes by a multipole expansion up to order $\ell_{\max}$, reducing the overall computational costs from $O(N^2K^2)$ to $O(NK\ell_{\max}^2)$. Finally, we show that the averaged dynamical system is equivalent to $2N$ interacting unit  spin vectors and provide two time integrators: a second-order symplectic scheme and a fourth-order Lie-group Runge--Kutta method, both of which are straightforward to generalize to higher order.
These new simulations recover the diffusion coefficients of stellar eccentricities obtained through analytical calculations of the secular dynamics.
\end{abstract}

\section{Introduction}
\label{sec:Introduction}

Supermassive \BHs\ are ubiquitous in external galaxies~\citep{KormendyHo2013}, where their active feedback plays a critical role in regulating galaxy formation through cosmic time~\citep{HeckmanBest2014}. Yet, the details of their diet and their impact on the stellar cluster that surrounds them (the galactic nucleus) remain open and challenging questions. Indeed, galactic nuclei are among the densest stellar systems in the universe. Despite the high stellar density, the gravitational potential in galactic nuclei is dominated by the central supermassive \BH\@. As a result, stars follow quasi-Keplerian orbits, which get slowly distorted by the additional perturbations present in the system.

The steep potential well generated by the central supermassive \BH\ implies the existence of a wide range of dynamical timescales in the system, and the evolution of the stellar cluster involves numerous dynamical processes acting on radically different timescales~\citep{RauchTremaine1996, HopmanAlexander2006, Merritt2013, Alexander2017}. These successively include: (i) the dynamical time associated with the fast Keplerian motion; on timescales longer than this, the stellar orbits can be regarded as eccentric massive wires; (ii) the in-plane precession time of the Keplerian wires generated by the relativistic corrections from the \BH\ and the stellar mean potential; (iii) the vector resonant relaxation time~\citep*[see, e.g.,][]{KocsisTremaine2015, FouvryBarOrChavanis2019}, which, owing to non-spherical stellar fluctuations and the relativistic corrections induced by a spinning \BH\@, leads to the reshuffling of the orientations of the orbital planes of the wires; (iv) the scalar resonant relaxation time~\citep[see, e.g.,][]{RauchTremaine1996, BarOrAlexander2016, SridharTouma2016, BarOrFouvry2018}, during which resonant torques between the precessing wires lead to a diffusion of the wires' eccentricities; (v) the non-resonant relaxation time~\citep{BahcallWolf1976, LightmanShapiro1977, CohnKulsrud1978, BarOrAlexander2013}, during which localised two-body encounters between stars lead to the long-term relaxation of the stars' Keplerian energy, i.e., the wires' semi-major axes.

As a result of this wide range of dynamical times,
from a few years for the fast Keplerian motion
(even a few minutes for stars near the event horizon)
up to a Hubble time for non-resonant relaxation in the nucleus of the Milky Way, direct numerical simulations of these dynamical systems remain very challenging. These were first performed with grid methods~\citep[see, e.g.\@,][]{Jacobs+2001,Kazandjian+2013}; it is only recently that an effective direct simulation of a galactic nucleus with $N = 10^{6}$ stars has been presented~\citep{Panamarev2019}, and even in this case most of the stars lie outside the central \BH\@'s sphere of influence. Conversely, simulations in the very relativistic regime are still limited to a small number of particles $N \!\sim\! 10^{2-3}$, should they use direct $N$-body methods~\citep{Merritt2011} or effective ones~\citep*[see, e.g.,][]{Madigan2011, Hamers2014}.

To circumvent these intrinsic difficulties, one has to resort to additional assumptions. Traditionally, the secular approximation smears out the stars along their underlying fast Keplerian motion, in other words replaces stars with Keplerian wires. Describing the dynamics of the stellar cluster amounts then to describing the long-term evolution of the wires' orbital parameters. This is in particular at the heart of the Gau{\ss} method~\citep*{ToumaTremaine2009} which provides an efficient algorithm to compute the force between two such wires. Should one be interested in the process of vector resonant relaxation, i.e., the relaxation of the orientations of orbital planes, this same approach can be further leveraged to also average the wires' dynamics over their in-plane precession, replacing Keplerian wires with Keplerian annuli, offering new venues to perform numerical investigations of these long-term dynamics~\citep{KocsisTremaine2015}.

The main benefit from these approaches is that any explicit average over the fast orbital motion offers a reduction of the range of timescales in the system. These methods can then explore longer timescales, beyond the reach of naive direct methods. Yet, most of these approaches suffer from relying on the computation of all the individual forces between objects, i.e., these methods come with a numerical complexity scaling like $O (N^{2})$ with $N$ the total number of stars.

In the present paper, we show how a multipole expansion~\citep{Henon1964, Aarseth1967, Henon1973, vanAlbada1977, FryPeebles1980, Villumsen1982, White1983, McGlynn1984, Meiron2014, Dehnen2014} yields a numerical scheme that can integrate the secular dynamics of Keplerian wires with a complexity scaling like $O(N K \ellmax^{2})$, with $N$ the total number of stars, $K$ a parameter independent of $N$, and $\ellmax$ the maximum harmonics considered in the multipole expansion.
We also show that the system is equivalent to $2N$ classical spin vectors,
and devise time integration schemes that exactly comply
with this system's geometric constraints.

The present paper is organised as follows. In Section~\ref{sec:Hamiltonian}, we describe the Hamiltonian and obtain its orbit-average to account for the domination of the central \BH\@. In Section~\ref{sec:EqMotion}, we derive the equations of motion for the orbital elements of the Keplerian wires. Section~\ref{sec:forces} details our numerical approach of discretising averages over the Keplerian motions as sums over nodes, and shows how to utilize a multipole expansion of the Newtonian pairwise interactions. In Section~\ref{sec:IntegrationScheme}, we describe numerical time integration schemes appropriate for the system of $N$ Keplerian wires. In Section~\ref{sec:RR}, we illustrate these new numerical methods by measuring the diffusion coefficients of stellar eccentricities and comparing them with predictions from kinetic theory. In Section~\ref{sec:Discussion} we discuss limitations of the present approach and possible future improvements, and we conclude in Section~\ref{sec:Conclusion}.

\section{The orbit-averaged Hamiltonian}
\label{sec:Hamiltonian}

We consider $N$ stars with masses $m_{i}$ and positions $\vec{R}_{i}$, orbiting a supermassive \BH\ of mass $\MBH$ at location $\vec{R}_{\bullet}$. The total Hamiltonian of this system is
\begin{align}
    H&= \frac{\vec{P}_{\bullet}^{2}}{2 \MBH} + \sum_{i=1}^{N}       \left[\frac{\vec{P}_{i}^{2}}{2 m_{i}} - \frac{G \MBH m_{i}}{|\vec{R}_{i} - \vec{R}_{\bullet}|} \right] + \sum_{i = 1}^{N} m_{i} \Phi_{\rel}^{i}
    \nonumber \\
    &- \, \sum_{i < j}^{N} \frac{G m_{i} m_{\!j}}{|\vec{R}_{i} - \vec{R}_{\!j}|} ,
\label{total_H}
\end{align}
with the canonical momenta $\vec{P}_{\bullet} \!=\! M_{\bullet} \dot{\vec{R}}_{\bullet}$ and $\vec{P}_{i} \!=\! m_{i} \dot{\vec{R}}_{i}$. In this equation, the potential contribution $\Phi_{\rel}^{i}$ accounts for the (conservative) relativistic corrections induced by the central \BH\@\footnote{The notation $\Phi_{\rel}$ is somewhat misleading, as the relativistic changes given in Appendix~\ref{sec:RelativisticPrecessions} cannot rigorously be derived from a potential. We nonetheless use this notation for convenience.}, i.e., the Schwarzschild and Lense--Thirring precessions~\citep[see, e.g\@,][]{Merritt2013}, whose detailed expressions are given in Appendix~\ref{sec:RelativisticPrecessions}.

\subsection{Democratic coordinates}
\label{sec:DemoCoordinates}

In order to emphasise the dominant influence of the central \BH\ on the system's dynamics, we rewrite equation~\eqref{total_H} using democratic coordinates centered on the \BH\@~\citep{Duncan1998} and their canonical momenta
\begin{subequations}
\begin{align}
\vec{r}_{\bullet} &= \frac{1}{\Mtot} \bigg[ \MBH \vec{R}_{\bullet} + \sum_{i = 1}^{N} m_{i} \vec{R}_{i} \bigg] , \quad \vec{r}_{i} = \vec{R}_{i} - \vec{R}_{\bullet} ,
\\
\vec{p}_{\bullet} &= \vec{P}_{\bullet} + \sum_{i = 1}^{N} \vec{P}_{i} , \quad  \vec{p}_{i} = \vec{P}_{i} - \frac{m_{i}}{\Mtot} \bigg[ \vec{P}_{\bullet} + \sum_{i = 1}^{N} \vec{P}_{\!i} \bigg] ,
\label{democratic_coordinates}
\end{align}
\end{subequations}
where we introduced the total mass $\Mtot =\MBH + \Mstar$ with $\Mstar = \sum_{i=1}^{N} m_{i}$ the total stellar mass. Thus, $\vec{p}_{\bullet}$ is just the total momentum, which we set to zero 
without loss of generality, so that $\vec{p}_i=\vec{P}_i$ is the barycentric momentum of the $i$th star. Following this change of coordinates, the Hamiltonian~\eqref{total_H} becomes
\begin{align}
H = & \, \sum_{i = 1}^{N} \left[ \frac{\vec{p}_{i}^{2}}{2 m_{i}} - \frac{G \MBH m_{i}}{|\vec{r}_{i}|} \right] + \sum_{i = 1}^{N} m_{i} \, \Phi_{\rel}^{i} - \sum_{i < j}^{N} \frac{G m_{i} m_{\!j}}{|\vec{r}_{i} - \vec{r}_{\!j}|}
\nonumber
\\
+ & \, \frac{1}{2 \MBH} \left[ \sum_{i = 1}^{N} \vec{p}_{i} \right]^{2}.
\label{total_H_demo}
\end{align}
The first term is the sum of $N$ independent Kepler Hamiltonians, the second term is associated with the relativistic corrections to stellar motion induced by the central \BH\@, the third term captures the pairwise interactions between the stars, and finally the last term is the kinetic energy of the central \BH. As we describe below,  after averaging over the fast Keplerian motions, the first and last term become irrelevant constants.

\subsection{Orbital Elements}
\label{sec:OrbitalElements}

In order to describe the Keplerian dynamics imposed by the central \BH\@, we transform the $(\vec{r} , \vec{p})$ coordinates to the Delaunay variables $(M , \omega , \Omega , \Lambda , L , L_{z}) $~\citep{BinneyTremaine2008} for each star. In this notation, the dynamical angles $M,\,\omega$, and $\Omega$ are, respectively, the mean anomaly, the argument of pericentre, and the longitude of the ascending node. The associated actions are
\begin{equation}
\Lambda = m \sqrt{G \MBH a} ,\quad
L = \Lambda \sqrt{1 - e^{2}} ,\quad
L_{z} = L \cos I ,
\label{def_L}
\end{equation}
where $\Lambda$ is the circular angular momentum of an orbit with the same energy or same semi-major axis $a$, $L$ the magnitude of the angular momentum vector, and $L_{z}$ its projection onto the $z$-axis, while $e$ and $I$ denote, respectively, the eccentricity and inclination of the orbit. Kepler's equation
\begin{equation}
\label{def_E}
M = E - e \sin E,
\end{equation}
introduces the eccentric anomaly $E$, which relates to the orbital radius, i.e., the distance from the central \BH\@, via $r=a(1-e \cos E)$.

Following this change of variables, equation~\eqref{total_H_demo} becomes
\begin{align}
H = - & \, \sum_{i = 1}^{N} \frac{m_{i}^{3} (G \MBH)^{2}}{2\Lambda_i^{2}} 
+ H_{\rel} + H_{\pw}
+ \frac{1}{2 \MBH} \left[ \sum_{i = 1}^{N} \vec{p}_{i} \right]^{2}.
\label{total_H_orbital}
\end{align}
with
\begin{align}
    \label{Hrel,Hw}
    H_{\rel} &\equiv
    \sum_{i=1}^{N} m_{i} {\Phi}_{\rel}^{i},&
    H_{\pw} &\equiv
    -\sum_{i<j}^{N}     \frac{Gm_{i}m_{\!j}}{|\vec{r}_{i} - \vec{r}_{\!j} |}.
\end{align}

\subsection{Orbit average}
\label{sec:OrbitAverage}

The first term in the Hamiltonian~\eqref{total_H_orbital} describes $N$ Keplerian orbits around the \BH\@, and only depends on a single set of actions, $\Lambda_i$. As a result, under this Hamiltonian, all variables but $M_{i}$ are conserved, and the mean anomaly $M$ evolves with the Keplerian orbital frequency
\begin{equation}
\Omega_{\Kep} (a) = \sqrt{\frac{G \MBH}{a^{3}}} .
\label{def_OmegaKep}
\end{equation}
If the \BH\ is supermassive, i.e., if one has $\MBH \gg \Mstar$, the dynamics is dominated by this fast Keplerian motion and the evolution of all other variables is much slower. Therefore, we average the stellar dynamics over these fast orbital motions, through the so-called secular approximation, to obtain the orbit-averaged Hamiltonian
\begin{align}
\label{def_oH}
\av{}{H} \big( \{\omega,\Omega,\Lambda,L,L_{z}\}_i \big)
    \equiv \avint{1}\dots\avint{N} H.
\end{align}
Here and in the remainder of this paper, we use the notation $\av{}{\cdot}$ for an average over all unperturbed stellar orbits.

Since $\av{}{H}$ is independent of the $M_{i}$, the associated actions $\Lambda_i$ are conserved and the first term of equation~\eqref{total_H_orbital} averages to a constant. Upon expanding, the last term of equation~\eqref{total_H_orbital} consists of the orbit-averaged stellar kinetic energies $\big\langle\vec{p}_{i}^{2}\big\rangle$ and products of the orbit-averaged stellar momenta $\av{}{\vec{p}_{i}}$. In our barycentric frame, $\av{}{\vec{p}_{i}}=0$ and $\big\langle\vec{p}_{i}^{2}\big\rangle=\big(m_i^2GM_\bullet/\Lambda_i\big)^2$ by virtue of the virial theorem. The constant terms depending only on the $\Lambda_i$ do not induce any dynamics and can be omitted, so the orbit-averaged Hamiltonian~\eqref{def_oH} finally becomes
\begin{align}
    \label{total_oH}
    \av{}{H}=\av{}{H_{\rel}}+\av{}{H_{\pw}}
\end{align}

\section{The equations of motion}
\label{sec:EqMotion}

The orbit-averaged Hamiltonian~\eqref{total_oH} describes the dynamics of $N$ gravitationally coupled Keplerian wires subject to relativistic precession. Each wire is characterised by the five quantities $\omega,\,\Omega,\, \Lambda,\,L,\,L_{z}$, of which $\Lambda$ is conserved through the orbit-averaged dynamics. Of course, the evolution equations for the four other coordinates can be obtained from Hamilton's canonical equations of motion, which involves obtaining the corresponding derivatives of $\vec{r}(\omega,\Omega,L,L_{z})$, at fixed $(M , \Lambda)$. Such calculations are rather involved, and can become degenerate, e.g., at $I=0$ (equatorial orbits), $e=0$ (circular orbits), or $e=1$ (radial orbits). An equivalent alternative is to work directly with the forces acting on the wires, as we will now pursue.

Rather than integrating the motion w.r.t.\ the orbital elements, we keep track of the wire dynamics through the dimensionless vectors
\begin{equation}
    \vec{h} = \frac{\vec{r} \times \vec{p}}{\Lambda},\qquad
    \vec{e} = \frac{\vec{p} \times (\vec{r} \times \vec{p})}{m^{2} G \MBH} - \uvec{r} ,
\label{def_bj_be_instant}
\end{equation}
where a hat denotes a unit vector as usual. Here, $\vec{h}$ is the angular momentum scaled to the circular angular momentum at the same energy and $\vec{e}$ is the eccentricity vector, which points in the direction of pericentre and has magnitude $|\vec{e}|=e$.

While only the four orbital elements $\omega,\,\Omega,\,L,\,L_{z}$ evolve, the two vectors $\vec{h},\,\vec{e}$ have six dynamical variables in total. The two associated degeneracies are captured by the two identities
\begin{equation}
\label{constraints_bj_be}
\vec{h} \cdot \vec{e} = 0,\qquad
\vec{h}^{2} + \vec{e}^{2} = 1 .
\end{equation}
In terms of these vectors, position and momentum are
\begin{subequations}
\label{exp_br_bp}
\begin{align}
\label{exp_br}
\vec{r} & \, = a \, \big( \cos E - e \big) \, \uvec{e} + a \sin E \; \vec{h} \times \uvec{e} ,
\\
\label{exp_bp}
\vec{p} & \, = - \frac{\Lambda}{r} \, \sin E \; \uvec{e} + \frac{\Lambda}{r} \, \cos E \; \vec{h} \times \uvec{e} .
\end{align}
\end{subequations}

\subsection{Orbit-averaged rates of change}
Using equations~\eqref{def_bj_be_instant} one can compute the time derivatives of the vectors $\vec{h}$ and $\vec{e}$ from the derivatives of the canonical variables $\vec{r}$ and $\vec{p}$.
The combined force from all other wires at position $\vec{r}_i$ on wire $i$ is
\begin{align}
    \vec{\mathcal{F}}_{\!i}(\vec{r}_i)
    &=-\frac{\partial \av{\backslash i}{H_{\pw}}}{\partial \vec{r}_{i}} = \sum_{j\neq i} G m_{i} m_{\!j} \,
    \av{\backslash i}{\vec{F}(\vec{r}_i-\vec{r}_{\!j})},
\label{def_ba}
\end{align}
where $\av{\backslash i}{\cdot}$ denotes an average over all orbits \emph{except} $i$ and $\vec{F}(\vec{r}) \equiv -\partial\varphi/\partial\vec{r}$ with $\varphi(\vec{r})\equiv-|\vec{r}|^{-1}$. At each value of the mean anomaly $M_i$ along the orbit, the force $\vec{\mathcal{F}}_{\!i}$ induces the local changes\footnote{The term in $\uvec{r}$ in equation~\eqref{def_bj_be_instant} does not contribute to the dynamics.} 
\begin{subequations}
\label{eqs:local:dt}
\begin{align}
    \label{eq:local:dh}
\dot{\vec{h}}_{i,\mathrm{local}}(M_i) &= \frac{1}{\Lambda_{i}}
    \vec{r}_{i} \times \vec{\mathcal{F}}_{\!i}
    - \frac{\dot{\Lambda}_i}{\Lambda_i} \vec{h}_i,
\\
    \label{eq:local:de}
\dot{\vec{e}}_{i,\mathrm{local}}(M_i) &= \frac{a_i}{\Lambda_i^2}
    \left[\vec{\mathcal{F}}_{\!i} \times (\vec{r}_{i} \times \vec{p}_{i}) + \vec{p}_{i} \times (\vec{r}_{i} \times \vec{\mathcal{F}}_{\!i})\right].
\end{align}
\end{subequations}
In order to obtain the rates of change $\dot{\vec{h}}_i$ and $\dot{\vec{e}}_i$, these local changes are averaged over the unperturbed Keplerian orbit $i$. The last term in equation~\eqref{eq:local:dh} averages to zero for conservative forces and we obtain
\begin{subequations}
\label{calc_der_time_II}
\begin{align}
    \dot{\vec{h}}_i &= \frac{1}{\Lambda_i} \sum_{j \neq i} G m_im_{\!j} \av{}{\vec{r}_{i} \times \vec{F}_{\!i\!j}},
\label{calc_der_time_II:j}
\\
\label{calc_der_time_II:e}
    \dot{\vec{e}}_i &= \frac{a_i}{\Lambda^2_i} \sum_{j \neq i} G m_im_{\!j} \av{}{
    \vec{F}_{\!i\!j} \times (\vec{r}_{i} \times \vec{p}_{i}) + \vec{p}_{i} \times (\vec{r}_{i} \times \vec{F}_{\!i\!j}) }.
\end{align}
\end{subequations}
Here, we used the shorthand $\vec{F}_{\!i\!j}\equiv\vec{F}(\vec{r}_i-\vec{r}_{\!j})$ and the averages are over all orbits \emph{including} $i$.

\subsection{Keeping all conservation laws}
The expressions~\eqref{calc_der_time_II} satisfy the constraints~\eqref{constraints_bj_be} for any conservative field $\vec{F}_{\!i\!j}$~\citep{ToumaTremaine2009}. 
Gau{\ss} showed that the average over orbit $j$ could be expressed in closed form. However, the double average over two interacting Keplerian wires $i$ and $j$ cannot be expressed in closed form but rather requires numerical treatment. The resulting numerical average will inevitably carry a small error with the consequence that the constraints~\eqref{constraints_bj_be} are no longer exactly honoured. While in general small numerical errors in conserved quantities are not a serious problem, this particular situation is awkward, since the conditions~\eqref{constraints_bj_be} are essential for the interpretation of $\vec{h}_i$ and $\vec{e}_i$ as a description of the wires. It is therefore important to construct numerical expressions for $\dot{\vec{h}}_i$ and $\dot{\vec{e}}_i$ that, despite their discretisation errors, keep the constraints~\eqref{constraints_bj_be} valid to machine precision.

In order to achieve that goal, we note that equations~\eqref{calc_der_time_II} are equivalent to  \citeauthor{Milankovitch1939}'s (\citeyear{Milankovitch1939}) relations \citep{TremaineTouma2009,Rosengren2014}
\begin{subequations}
\label{eq:Milankovitch}
\begin{align}
\label{Milan_h}
\dot{\vec{h}}_{i} {} & =  - \frac{1}{\Lambda_{i}} \left( \vec{h}_{i} \times \frac{\partial \av{}{H_{\star}} }{\partial \vec{h}_{i}} + \vec{e}_{i} \times \frac{\partial \av{}{H_{\star}}}{\partial \vec{e}_{i}} \right),
\\
\label{Milan_e}
\dot{\vec{e}}_{i} {} & = - \frac{1}{\Lambda_{i}} \left( \vec{h}_{i} \times \frac{\partial \av{}{H_{\star}}}{\partial \vec{e}_{i}} + \vec{e}_{i} \times \frac{\partial \av{}{H_{\star}}}{\partial \vec{h}_{i}} \right).
\end{align}
\end{subequations}
Here, the averaged Hamiltonian $\av{}{H_{\star}}$ is expressed as a function of the vectors $\vec{h}_i,\,\vec{e}_i$ (and the constants $\Lambda_i$) from all wires. 

These relations are equivalent to the canonical equations of motion, but much more useful. First, the vectors $\vec{h}$ and $\vec{e}$ are well-defined for all orbits, even those with zero eccentricity, unit eccentricity, and zero inclination. Second, they have the beautiful property that the identities~\eqref{constraints_bj_be} and the conservation of total energy
are explicitly satisfied for any form of $\av{}{H_\star}$, i.e.,
\begin{subequations}
\label{eq:cons}
\begin{align}
\label{eq:cons_norm}
\frac{\rd \, (\vec{h}_{i}^{2} + \vec{e}_{i}^{2})}{\rd t} &= \dot{\vec{h}}_{i} \cdot \vec{h}_{i} + \dot{\vec{e}}_{i} \cdot \vec{e}_{i} = 0,
\\
\label{eq:cons_dot}
\frac{\rd \, (\vec{h}_{i} \cdot \vec{e}_{i})}{\rd t} &= \dot{\vec{h}}_{i} \cdot \vec{e}_{i} + \dot{\vec{e}}_{i} \cdot \vec{h}_{i} = 0,
\\
\label{eq:cons_Etot}
\frac{\rd\av{}{H_\star}}{\rd t} &= \sum_i 
\frac{\partial \av{}{H_{\star}} }{\partial \vec{h}_{i}} \cdot\dot{\vec{h}}_i + \frac{\partial \av{}{H_{\star}} }{\partial \vec{e}_{i}} \cdot\dot{\vec{e}}_i = 0.
\end{align}
\end{subequations}
These properties suggest an alternative, fully conservative approach to numerically computing the rates $\dot{\vec{h}}_i$ and $\dot{\vec{e}}_i$: instead of following established practice of discretising the orbit averages in equations~\eqref{calc_der_time_II}, we first discretise the orbit average for $\av{}{H_\star}$ and subsequently use Milankovitch's relations~\eqref{eq:Milankovitch} to obtain $\dot{\vec{h}}_i$ and $\dot{\vec{e}}_i$\footnote{This method for deriving discretised equations of motion is analogous to conservative formulations of Smoothed Particle Hydrodynamics (SPH), where instead of discretising the Euler equation directly one discretises the fluid Lagrangian, such that the Euler-Lagrange equations of motion give rise to a discretised form of the Euler equation that honours all conservation laws \citep{SpringelHernquist2002}.}, as detailed in Section~\ref{sec:Nodes}.

\subsection{Reformulation as a spin system}
Following~\cite{Klein1924}, we introduce the vectors
\begin{equation}
    \label{def_b}
    \vec{b}_{i+} = \vec{h}_i + \vec{e}_i
    \quad\text{and}\quad
    \vec{b}_{i-} = \vec{h}_i - \vec{e}_i ;
\end{equation}
then the identities~\eqref{constraints_bj_be} become
\begin{equation}
    \label{constraints_bp_bm}
    | \vec{b}_{i\pm} | = 1.
\end{equation}
Thus, the system of $N$ wires is fully described by $2N$ independent unit vectors $\vec{b} = \vec{b}_{i\pm}$, in other words it is a classical spin system. The dynamics of these vectors is simply

\begin{equation}
    \label{vel_b}
    \dot{\vec{b}} = 
    \dvec{h} \pm \dvec{e}.
\end{equation}
Since the vectors $\vec{b}$ remain on the unit sphere, this equation can be expressed as a precession
\begin{equation}
    \label{prec_b}
    \dvec{b} = \vec{\Omega} \times \vec{b}.
\end{equation}
The precession vectors $\vec{\Omega}$ are not uniquely determined through equation~\eqref{vel_b}, since a component parallel to $\vec{b}$ does not affect $\dvec{b}$ (one would need to know $\ddot{\vec{b}}$ to construct such a component). The most conservative choice for the precession vectors---in the sense that $|\vec{\Omega}|^2$ is minimized---is therefore
\begin{equation}
    \label{eq:Omega}
    \vec{\Omega} = \vec{b} \times \dvec{b},
\end{equation}
for which $\vec{\Omega}\cdot\vec{b}=0$, so that $\vec{b}$ moves along a great circle for constant $\vec{\Omega}$.

Milankovitch's equations~\eqref{eq:Milankovitch}, when re-expressed in terms of the vectors $\vec{b}$, read
\begin{equation}
    \label{eq:Milankovitch:b}
    \dvec{b} = \frac{2}{\Lambda} \frac{\partial\av{}{H_\star}}{\partial\vec{b}} \times \vec{b}.
\end{equation}
These have the standard form for the equations of motion of classical spin systems. Comparing with equation~\eqref{prec_b}, one might identify  $2\Lambda^{-1}\partial\!\av{}{H_\star}\!/\partial\vec{b}$ with $\vec{\Omega}$. However, the gradient of the Hamiltonian is not unique, since owing to the constraints~\eqref{constraints_bp_bm}, $\av{}{H_\star}$ is only determined up to additive terms of the form $(\vec{b}^2-1) \tilde{H}$ with an arbitrary function $\tilde{H}$. Such terms alter the gradient, but 
have no effect on $\dvec{b}$, reflecting the gauge invariance of equation~\eqref{eq:Milankovitch:b}.

We are now set to compute the rates of change for each wire (Section~\ref{sec:forces}), and to perform the time-integration of their evolution (Section~\ref{sec:IntegrationScheme}). Throughout the coming sections, we will test our algorithm in three systems, namely a simple analytical ``Pair'' Hamiltonian of two oscillating wires, Kozai--Lidov oscillations of $N=2$ stars (whose orbits do not overlap radially) around a central \BH\@, and a $N = 10^{4}$ stellar cluster mimicking Sgr A*. We refer to Appendices~\ref{sec:Pair}, \ref{sec:KozaiLidov}, and~\ref{sec:PhysicalModel} for the detailed description of these setups.

\section{Calculating the rates of change}
\label{sec:forces}

Our computed rates of change $\dot{\vec{h}}_i$ and $\dot{\vec{e}}_i$ differ from the continuous forms~\eqref{calc_der_time_II} or~\eqref{eq:Milankovitch} first by estimating the averaging integrals by discrete sums (detailed in Section~\ref{sec:Nodes}), and second by approximating the forces using spherical harmonics (detailed in Section~\ref{sec:MultiPoleExpansion}).
We also refer to Appendix~\ref{sec:RelativisticPrecessions} for explicit expressions of the contributions to $\dot{\vec{h}}$ and $\dot{\vec{e}}$ from $\av{}{H_{\rel}}$,
namely the Schwarzschild and Lense--Thirring relativistic precessions.

\subsection{Discretised rates of change}
\label{sec:Nodes}
In order to numerically calculate the orbit averages, we approximate them by discrete midpoint sums over $K$ positions along each orbit\footnote{
    Adopting a different value of $K$ for each wire, e.g., depending on eccentricity, is a straightforward extension.}:
\begin{equation}
	\label{discrete}
    m_i \oint \frac{\rd M_i}{2\pi} \, f(M_i) \;\to\;
    \sum_{k=1}^K \mu_{ik}\, f(M_{ik})
\end{equation}
for any function $f(M)$. In other words, each eccentric wire of mass $m_i$ is replaced by $K$ \emph{nodes} of masses $\mu_{ik}$. Placing the nodes' mean anomalies $M_{ik}$ uniformly \citep{GurkanHopman2007} is a bad idea, not only because it requires solving Kepler's equation~\eqref{def_E} for each node, but also because it only poorly samples the pericentric passage, in particular for large eccentricity, and hence produces a slow convergence with $K$. Instead, we place the nodes uniformly in eccentric anomaly:
\begin{equation}
	\label{node_placement}
    M_{ik} = E_k - e_i \sin E_k,\qquad
    E_k = (k-\tfrac12) \Delta E
\end{equation}
with $\Delta E=2\pi/K$. With this choice, the nodes are placed symmetrically w.r.t.\ pericentre but there is no node exactly at that position. To complete the discretisation~\eqref{discrete}, we specify the node masses
\begin{align}
    \label{node:mass}
    \mu_{ik} = \frac{m_i}{2\pi} \frac{\rd M_i}{\rd E} \Delta E
             = \frac{m_i}{K} (1-e_i\cos E_k).
\end{align}
With these specifications, the discretised potential energy of node $k$ on wire $i$ becomes
\begin{equation}
\label{def_phik}
\phi_{ik} = \mu_{ik}\psi_{ik},\qquad
\psi_{ik} = \sum_{j \neq i}^N \sum_{l=1}^K G  \mu_{jl}\,\varphi(\vec{r}_{ik} - \vec{r}_{jl}),
\end{equation}
such that the total interaction energy between the wires is computed as
\begin{equation}
    \av{}{H_{\star}} = \frac12 \sum_{i=1}^N \sum_{k=1}^K \phi_{ik}.
    \label{Htot}
\end{equation}
This expression depends on the $\vec{h}_i$ and $\vec{e}_i$ through the node positions $\vec{r}_{ik}$ but also through their masses $\mu_{ik}$. The latter depend on the eccentricities $e_i$ because of our sampling of the nodes in eccentric rather than mean anomaly. The derivatives of $\av{}{H_{\star}}$ resulting from this dependence through $\mu_{ik}$ induce via Milankovitch's relations~\eqref{eq:Milankovitch} the rates of change (see Appendix~\ref{sec:EOM:disc})
\begin{subequations}
\label{EOM_mu}
\begin{align}
\label{dhdt_mu}
\dot{\vec{h}}_{i} \big|_{\mu} {} &= 0 ,
\\
\label{dedt_mu}
\dot{\vec{e}}_{i} \big|_{\mu} {} &= \frac{1}{\Lambda_{i}} \, \vec{h}_{i} \times \uvec{e}_{i} 
\sum_{k=1}^K \bar{\mu}_i\,\psi_{ik}\,\cos E_{k}.
\end{align}
\end{subequations}
where $\bar{\mu}_i\equiv m_i/K$ is the mean node mass on the wire.
These rates of change do not appear
in the traditional equations~\eqref{calc_der_time_II} based on forces.

The rates of change induced by the dependence of $\av{}{H_{\star}}$ on $\vec{h}_i$ and $\vec{e}_i$ through $\vec{r}_{ik}$ are (see Appendix~\ref{sec:EOM:disc})
\begin{subequations}
\label{EOM_r}
\begin{align}
\label{dhdt_r}
\dot{\vec{h}}_{i} \big|_{\vec{r}} {} & = \frac{a_{i}}{\Lambda_{i}} \big[ \uvec{e}_{i} \times \big( \vec{C}_{i} - e_{i} \vec{F}_{\!i} \big) + \big( \vec{h}_{i} \times \uvec{e}_{i} \big) \times \vec{S}_{i} \big] ,
\\
\label{dedt_r}
\dot{\vec{e}}_{i} \big|_{\vec{r}} {} & = \frac{a_{i}}{\Lambda_{i}} \bigg[ \vec{F}_{\!i} \times \vec{h}_{i} - e_{i} \vec{S}_{i} + \frac{1}{e_{i}} \big[ \vec{S}_{i} \cdot \uvec{e}_{i} - \vec{C}_{i} \cdot \big( \vec{h}_{i} \times \uvec{e}_{i} \big) \big] \uvec{e}_{i} \bigg]
\end{align}
\end{subequations}
with $\vec{F}_{\!i} \equiv \sum_{k} \vec{f}_{\!ik} $,
$\vec{C}_{i} \equiv \sum_{k} \vec{f}_{\!ik}\cos E_{k}  $,
and $\vec{S}_{i} \equiv \sum_{k} \vec{f}_{\!ik}\sin E_{k} $.
Here,
\begin{align}
\label{def_fik}
\vec{f}_{\!ik} {} & = -\frac{\partial \phi_{ik}}{\partial\vec{r}_{ik}} = \sum_{j \neq i}^N \sum_{l=1}^K G \mu_{ik} \, \mu_{jl} \, \vec{F}(\vec{r}_{ik} - \vec{r}_{jl})
\end{align}
is the gravitational force acting on node $k$ of wire $i$ and generated by the nodes of all other wires. Upon close inspection
\begin{equation}
\label{dhdt_r_rewrite}
    \dot{\vec{h}}_i\big|_{\vec{r}} = 
    \frac1{\Lambda_i}\sum_{k=1}^K \vec{r}_{ik}\times\vec{f}_{\!ik},
\end{equation}
which is identical to the discretisation of equation~\eqref{calc_der_time_II:j} and implies conservation of total angular momentum $\vec{L}_{\mathrm{tot}}=\sum_i\Lambda_i\vec{h}_i$, as is expected from Noether's theorem (since our discretisation of the wire averages remains invariant under spatial rotations).
Also note that the discretisation of equation~\eqref{calc_der_time_II:e} clearly differs from equation~\eqref{dedt_r}.
The advantage of equation~\eqref{dedt_r} is that it satisfies $\dot{\vec{h}}_i \cdot \vec{h}_i + \dot{\vec{e}}_i \cdot \vec{e}_i = 0$ and $\dot{\vec{h}}_i \cdot \vec{e}_i + \dot{\vec{e}}_i \cdot \vec{h}_i = 0$ exactly, as required.

We note from equations~\eqref{EOM_mu} and~\eqref{EOM_r} that the computation of $\dot{\vec{h}}_i$ and $\dot{\vec{e}}_i$ due to interactions with $N$ wires requires $O(N^2 K^2)$ computations. As shown below, this can be reduced to $O(N K \ell_{\rm max}^{2})$ when approximating  $\vec{F}(\vec{r}_{ik}-\vec{r}_{jl})$ by its expansion in spherical harmonics up to order $\ell_{\rm max}$. To this end, it is advantageous to split
$
    \vec{f}_{\!ik} = \vec{f}^{\all}_{\!ik} - \vec{f}^{\self}_{\!ik} 
$
into the contributions
\begin{subequations}
\begin{align}
    \label{fall:def}
    \vec{f}^{\all}_{\!ik} &=
    \sum_{\{j,l\}\neq\{i,k\}}^{N,K} G\mu_{ik}\,\mu_{\!jl}\,
    \vec{F}(\vec{r}_{ik}-\vec{r}_{\!jl}),
    \\
    \label{fself:def}
    \vec{f}^{\self}_{\!ik} &= \sum_{l\neq k}^K \, G\mu_{ik}\,\mu_{il}\,
    \vec{F}(\vec{r}_{ik}-\vec{r}_{il}),
\end{align}
\end{subequations}
due to all of the other nodes on all of the wires, and due to all of the other nodes on the same wire, respectively.

We stress that our algorithms
for $\vec{f}^{\all}_{ik}$ and $\vec{f}^{\self}_{ik}$ must compute the exact same wire self-gravity such that when computing the difference $\vec{f}_{\!ik} = \vec{f}^{\all}_{ik} - \vec{f}^{\self}_{ik}$ these erroneous non-physical contributions exactly cancel.
This is important, since the wire self-gravity can be substantial -- it diverges logarithmically with the number $K$ of nodes in the case of exact (unsoftened) Newtonian gravity.
Similarly, our algorithms must also ensure that a single Keplerian wire
is a stationary configuration of the orbit-averaged dynamics.

\subsection{Multipole Expansion}
\label{sec:MultiPoleExpansion}

In order to accelerate the computation of the $O(N^2K^2)$ pairwise node interactions, we utilise the expansion of the Newtonian interaction kernel $\varphi(\vec{r})=-|\vec{r}|^{-1}$ in spherical harmonics
\begin{align}
    \label{Legendre_Expansion}
   \displaystyle 
    \varphi(\vec{r}_{i} - \vec{r}_{\!j})
    &= -\sum_{\ell = 0}^{\infty}\sum_{m = - \ell}^{\ell}
    \begin{cases}
        U_\ell^m (\vec{r}_{i}) \, T_\ell^m (\vec{r}_{\!j})
        & \text{if $r_i<r_j$}, \\
        T_\ell^m (\vec{r}_{i}) \, U_\ell^m (\vec{r}_{\!j}) 
        & \text{if $r_i>r_j$}.
    \end{cases}
\end{align}
Here, we have defined real-valued upper and lower solid spherical harmonics, $U_\ell^m(\vec{r})$ and $T_\ell^m(\vec{r})$ respectively, which are described explicitly in Appendix~\ref{sec:SphericalHarmonics}. 
Note that there exist essentially two flavours of codes based on spherical harmonic expansions: (i) codes in which the radial forces are evaluated using a basis-function expansion~\citep[see, e.g.,][]{HernquistOstriker1992,Saha1993}; (ii) codes in which the radial forces associated with a given spherical harmonic are evaluated exactly~\citep[see, e.g.,][]{Villumsen1982} using formulas involving the usual factor $r_{\mathrm{min}}^{\ell}/r_{\mathrm{max}}^{\ell+1}$, as in equation~\eqref{Legendre_Expansion}, here absorbed into the definitions of $U_\ell^m$ and $T_\ell^m$. Our goal here is to follow this second avenue and tailor it to the case of Keplerian wires. The expansion~\eqref{Legendre_Expansion} is isotropic, even when truncated at order $\ellmax$, and hence, by virtue of Noether's theorem, will conserve the total angular momentum $\Ltot$ of the system. However, the expansion is not invariant under translations and therefore the total linear momentum is usually not conserved in spherical-harmonic codes~\citep[][\S2.9.4]{BinneyTremaine2008}. Fortunately, as already highlighted in equation~\eqref{def_oH}, Keplerian wires do not induce any force on the central object, which ensures, by design, the conservation of the system's total linear momentum.

Inserting the expansion~\eqref{Legendre_Expansion} into equation~\eqref{fall:def}, we have
\begin{align}
\label{Hall}
\vec{f}^{\all}_{ik} &= \sum_{j,l}^{N,K} \sum_{\ell,m} G\mu_{ik}\,\mu_{\!jl}
\begin{cases}
    \displaystyle U_\ell^m(\vec{r}_{\!jl})\, \vec{\nabla} T_\ell^m(\vec{r}_{ik})
    & \text{if $r_{\!jl} < r_{ik}$}
    \\
    \displaystyle T_\ell^m(\vec{r}_{\!jl})\, \vec{\nabla} U_\ell^m(\vec{r}_{ik})
    & \text{if $r_{\!jl} > r_{ik}$},
\end{cases}
\end{align}
with an analogous expression for $\vec{f}^{\self}_{ik}$. The important difference between this expression and equation~\eqref{fall:def} is that the dependence on the positions $\vec{r}_{ik}$ and $\vec{r}_{\!jl}$ has been factorized. Hence, the gradient terms depending on $\vec{r}_{ik}$ can be taken outside the sum over nodes after splitting it into an inner and outer part, giving
\begin{align}
    \label{fall:PQ}
    \vec{f}^{\all}_{ik} &= 
    \sum_{\ell,m} \mu_{ik}
        P_\ell^m(r_{ik}) \,\vec{\nabla} T_\ell^m (\vec{r}_{ik}) +
        \sum_{\ell,m} \mu_{ik}
        Q_\ell^m(r_{ik}) \,\vec{\nabla} U_\ell^m (\vec{r}_{ik}),
\end{align}
where
\begin{subequations}
\label{eqs:P,Q}
\begin{align}
    P_\ell^m(r) &= \sum_{j,l:\,r_{\!jl}<r} G\mu_{\!jl}\, U_\ell^m (\vec{r}_{\!jl}),
    \\
    Q_\ell^m(r) &= \sum_{j,l:\,r_{\!jl}>r} G\mu_{\!jl}\, T_\ell^m (\vec{r}_{\!jl})
\end{align}
\end{subequations}
are the multipoles of the distribution of all nodes inside and outside of radius $r$, respectively.

We note that the expansion from equation~\eqref{Legendre_Expansion} may also be used to define a cluster's truncated total energy, $\Etot (\ellmax,K)$, via equation~\eqref{total_oH}, which can be computed efficiently using the same algorithm as for the wire forces. In the limit $\ellmax \to + \infty$, i.e.\ in the absence of any multipole expansion, equation~\eqref{total_oH} provides us with the full total energy, $\Etotdirect$, which we can also compute via a direct $O(N^2 K^2)$ sum over all pairwise node-node couplings.

\subsection{Computing the wire forces}
\label{sec:calcWire}

We begin by sorting the nodes into ascending order of radius\footnote{For the self-gravity to be correctly removed, it is mandatory for the radius sortings used to compute $\vec{f}_{\alpha}^{\all}$ and $\vec{f}_{\alpha}^{\self}$ to be consistent.}. Since the nodes on each wire are already sorted when created, this costs only $O(NK\ln N)$ rather than $O(NK\ln(NK))$ operations and is completely subdominant in the total operation budget of our code.

For brevity we denote the pair of indices $i$ labeling the wire and $k$ labeling the node on the wire by $\alpha=\{i,k\}$, which we call the node index. Once the nodes are sorted in radius, the multipoles $P_\ell^m$ and $Q_\ell^m$ can be calculated by increasing and decreasing recurrence, respectively. The algorithm for the computation of $\vec{f}^{\all}_\alpha$ for all nodes is summarized in Table~\ref{tab:algorithm}.

\begin{table}
\begin{enumerate}
    \itemsep0pt\parsep0pt\topsep0pt\partopsep0pt\parskip0pt
    \item[1] sort all nodes in radius $r_\alpha$.
    \item[2] set $\vec{f}^{\all}_\alpha=0$ for all nodes $\alpha$ 
    \item[3] set $P_{\ell}^m=0$
    \item[4] for all nodes $\alpha$ in order of increasing radius $r_\alpha$:
    \begin{enumerate}
    \itemsep0pt\parsep0pt\topsep0pt\parskip2pt
        \item[4.1] $\vec{f}^{\all}_\alpha
            \,\leftarrow\, \vec{f}^{\all}_\alpha + \mu_\alpha\sum_{\ell,m} P_{\ell}^m\,\vec{\nabla}T_{\ell}^m(\vec{r}_\alpha)$,
        \item[4.2] $P_{\ell}^m \,\,\leftarrow\, P_{\ell}^m \, + G\mu_\alpha U_\ell^m(\vec{r}_\alpha)$,
    \end{enumerate}
    \item[5] set $Q_{\ell}^m=0$
    \item[6] for all nodes $\alpha$ in order of decreasing radius $r_\alpha$:
    \begin{enumerate}
    \itemsep0pt\parsep0pt\topsep0pt\parskip2pt
        \item[6.1] $\vec{f}^{\all}_\alpha
            \,\leftarrow\, \vec{f}^{\all}_\alpha + \mu_\alpha\sum_{\ell,m} Q_{\ell}^m\,\vec{\nabla}U_\ell^m(\vec{r}_\alpha)$,
        \item[6.2] $Q_{\ell}^m \,\leftarrow\, Q_{\ell}^m + G\mu_\alpha T_{\ell}^m(\vec{r}_\alpha)$;
    \end{enumerate}
\end{enumerate}
\caption{
The algorithm to compute $\vec{f}^{\all}_\alpha$ for all $NK$ nodes. The forces on node $\alpha$ due to all other nodes at smaller and large radii, respectively, are computed in steps 4.1 and 6.1 from the inner and outer multipoles $P_\ell^m$ and $Q_\ell^m$, which in turn are accumulated in steps 4.2 and 6.2.
\label{tab:algorithm}
}
\end{table}
The total operation count of this algorithm is $O(NK\ell_{\max}^2)$ with $\ell_{\max}$ the maximum harmonic order considered. The same algorithm, but restricted to the nodes from only one wire, calculates $\vec{f}^{\self}_\alpha$ for that wire and again requires $O(NK\ell_{\max}^2)$ operations for all wires.

\subsection{Convergence of the force calculations}
\label{sec:ConvergenceForces}

\begin{figure}
\begin{center}
\includegraphics[width=0.45\textwidth]{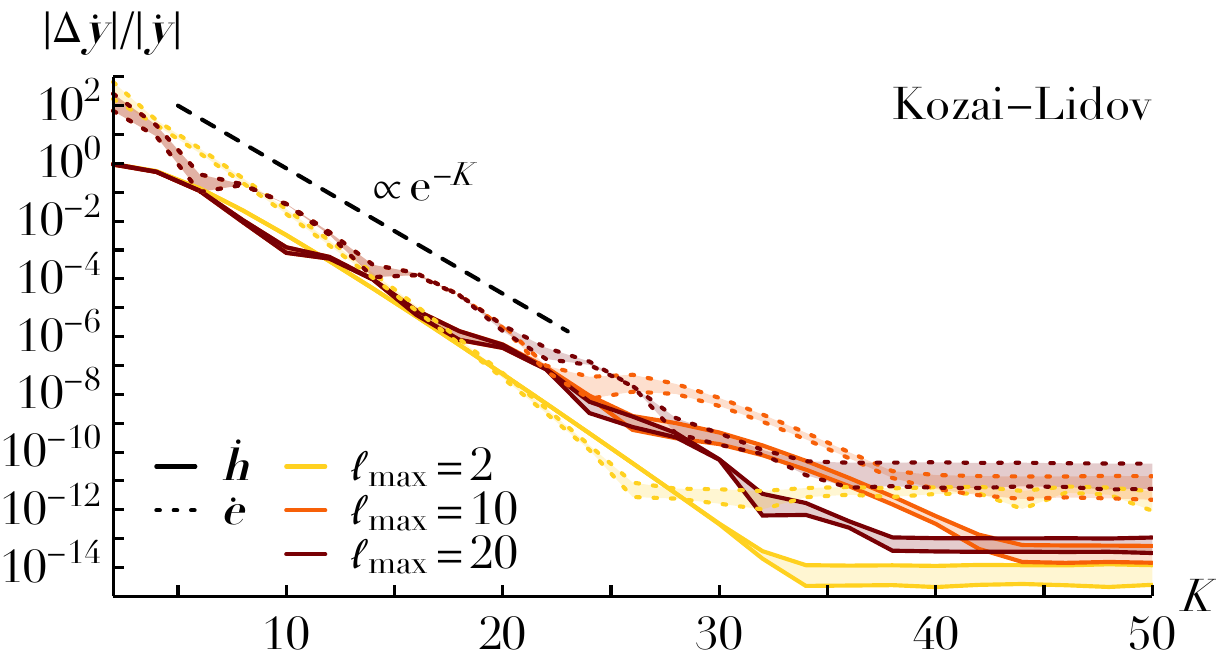}
\\
\includegraphics[width=0.45\textwidth]{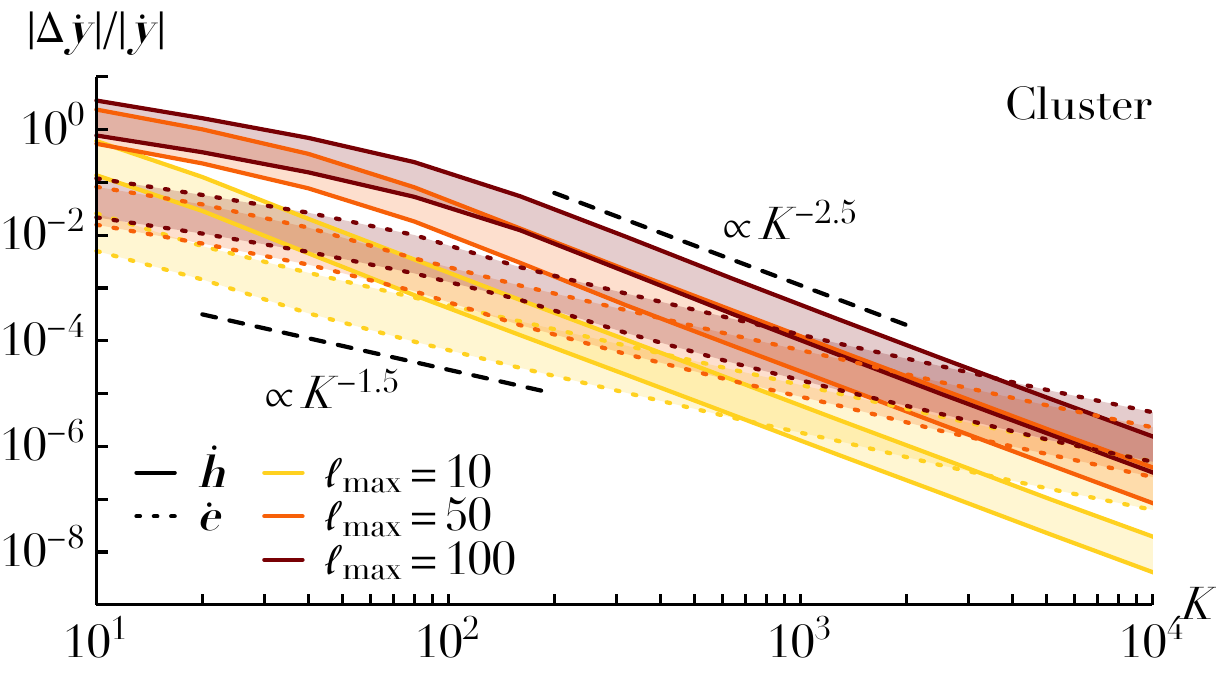}
\end{center}
\caption{Illustration of the relative errors in $\dot{\vec{h}}$ and $\dot{\vec{e}}$ as a function of the number of nodes $K$ for fixed multipole truncation order $\ellmax$, for Kozai--Lidov oscillations (Appendix~\ref{sec:KozaiLidov}) and an $N=10^{4}$ stellar cluster (Appendix~\ref{sec:PhysicalModel}). We used 500 independent realisations of the Kozai--Lidov system in the top panel, and a single cluster for the bottom panel. The colored regions correspond to the 16\% and 84\% levels. We compute the relative errors by comparison to calculations with $K=100$ for the top panel and $K=20\,480$ for the bottom panel.
\label{fig:errK}}
\end{figure}

In Figure~\ref{fig:errK}, we illustrate the dependence of the relative errors in both $\dot{\vec{h}}$ and $\dot{\vec{e}}$ as a function of the number of nodes per wire $K$.

For the Kozai--Lidov system, for a fixed value of $\ellmax$, we find that the error decreases exponentially as $\re^{-K}$. This is a direct consequence of the absence of any radial overlap between the two stellar orbits at play. In that case, the double orbit-average integral from equation~\eqref{def_oH} naturally splits into the product of two $1D$ integrals, which the multipole algorithm from equation~\eqref{eqs:P,Q} catches. Importantly, the integrands of each of these $1D$ integrals are $2\pi$-periodic w.r.t.\ the eccentric anomaly. In that case, the midpoint sampling from equation~\eqref{node_placement} ensures an exponential convergence w.r.t.\ the number of nodes $K$~\citep{Trefethen+2014}.

For the $N=10^{4}$ clusters, given that the orbits exhibit a wide range of eccentricities -- $0 \leq e \leq 0.99$, see Appendix~\ref{sec:PhysicalModel} -- radial overlaps are ubiquitous and the exponential convergence w.r.t.\ $K$ does not hold anymore. More precisely, for a fixed value of $\ellmax$, the error decreases as $K^{-2.5}$ for $\dot{\vec{h}}$ and as $K^{-1.5}$ for $\dot{\vec{e}}$. We believe that this difference in the rates of convergence stems from the fact that the discontinuity in equation~\eqref{Hall} is only in the radial force, which does not enter the equation for $\dot{\vec{h}}$.

\begin{figure}
\begin{center}
\includegraphics[width=0.45\textwidth]{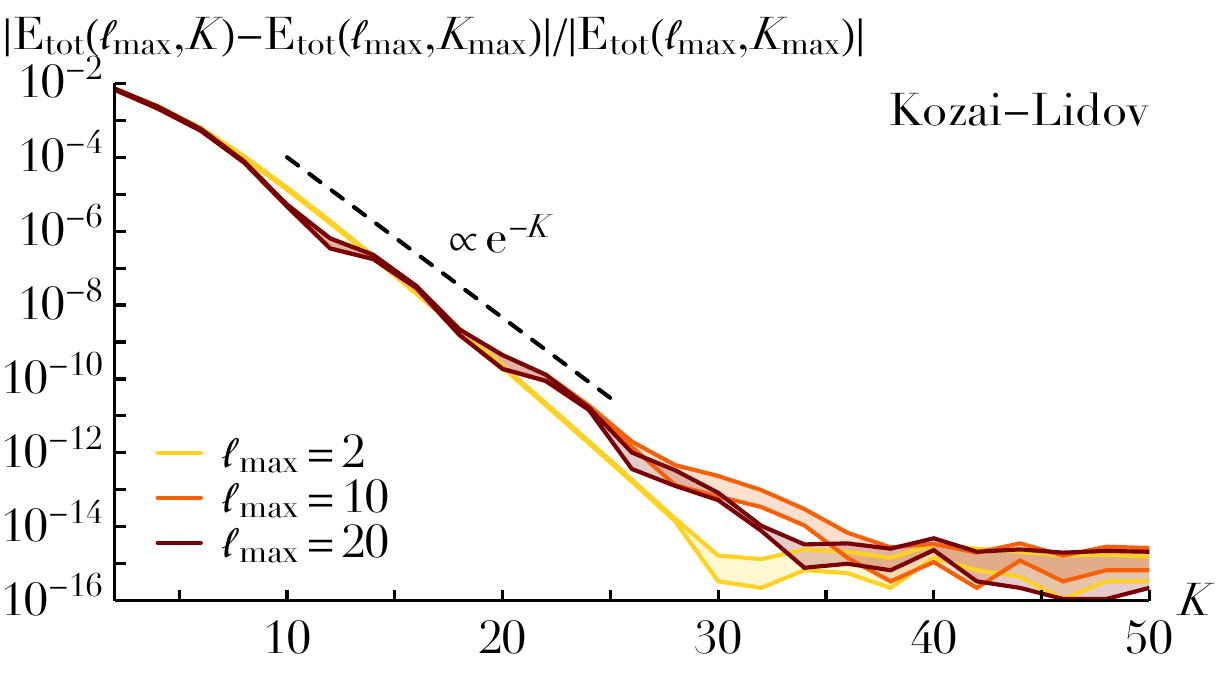}
\\
\includegraphics[width=0.45\textwidth]{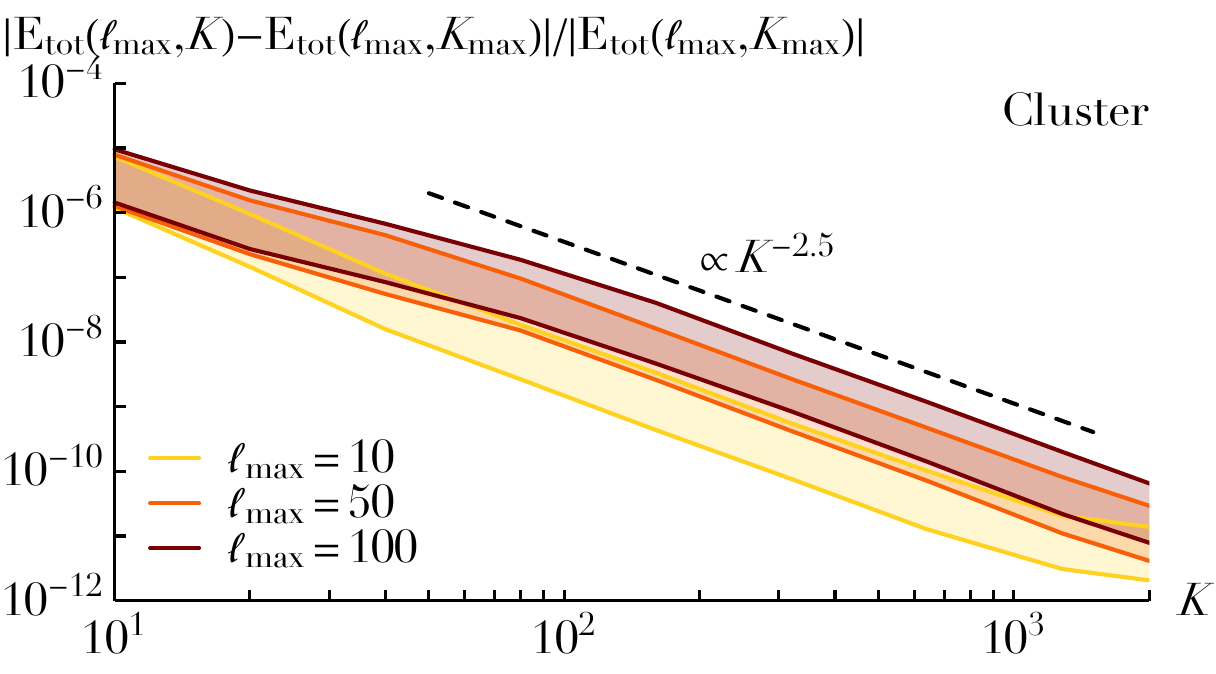}
\end{center}
\caption{Illustration of the relative error in the truncated total energy, $ \Etot (\ellmax,K)$,
as a function of the number of nodes $K$
for fixed multipole truncation order $\ellmax$,
w.r.t.\ to calculations with $\Kmax = 100$ for the top panel
and $\Kmax = 20\,480$ for the bottom panel.
In both panels, the colored regions correspond
to the 16\% and 84\% levels
over 500 independent realisations.
\label{fig:errEtot_K}}
\end{figure}

We reach similar conclusions in Figure~\ref{fig:errEtot_K},
where we present the error in the truncated total energy,
$\Etot (\ellmax,K)$ as a function of $K$.
For the radially non-overlapping Kozai--Lidov simulations,
it converges exponentially w.r.t.\ $K$,
while for the $N=10^{4}$ clusters,
it is found to converge like $K^{-2.5}$,
likely a consequence of the absence
of any discontinuities in the expansion
from equation~\eqref{Legendre_Expansion}.

\begin{figure}
\begin{center}
\includegraphics[width=0.45\textwidth]{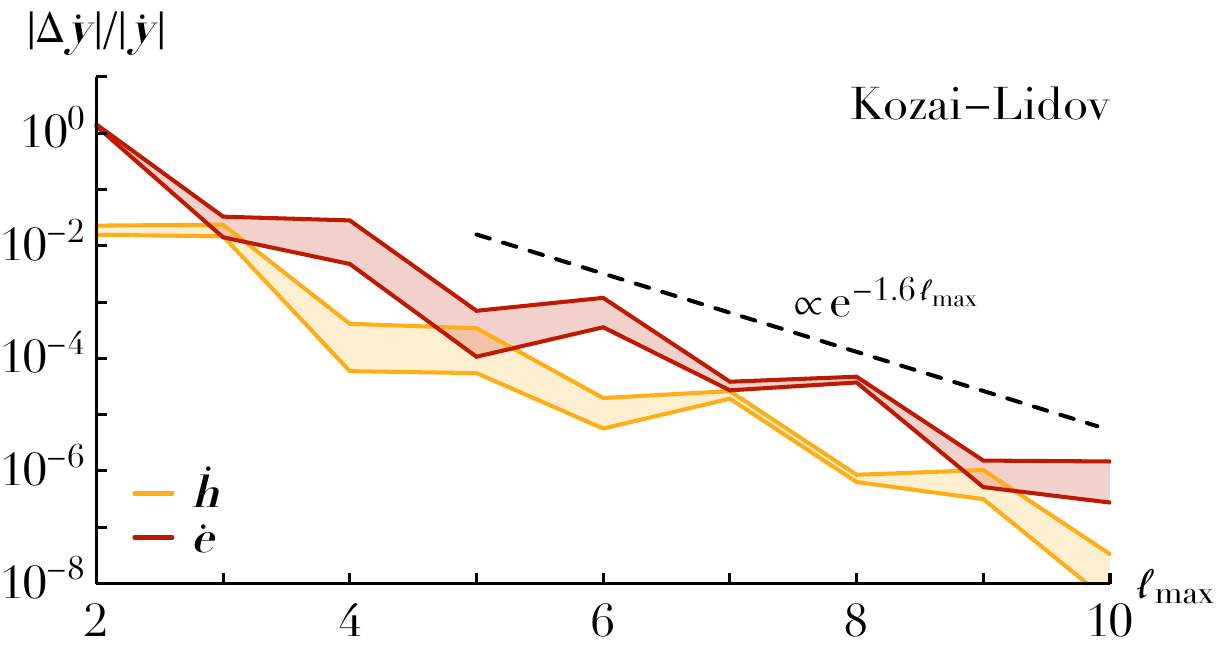}
\\
\includegraphics[width=0.45\textwidth]{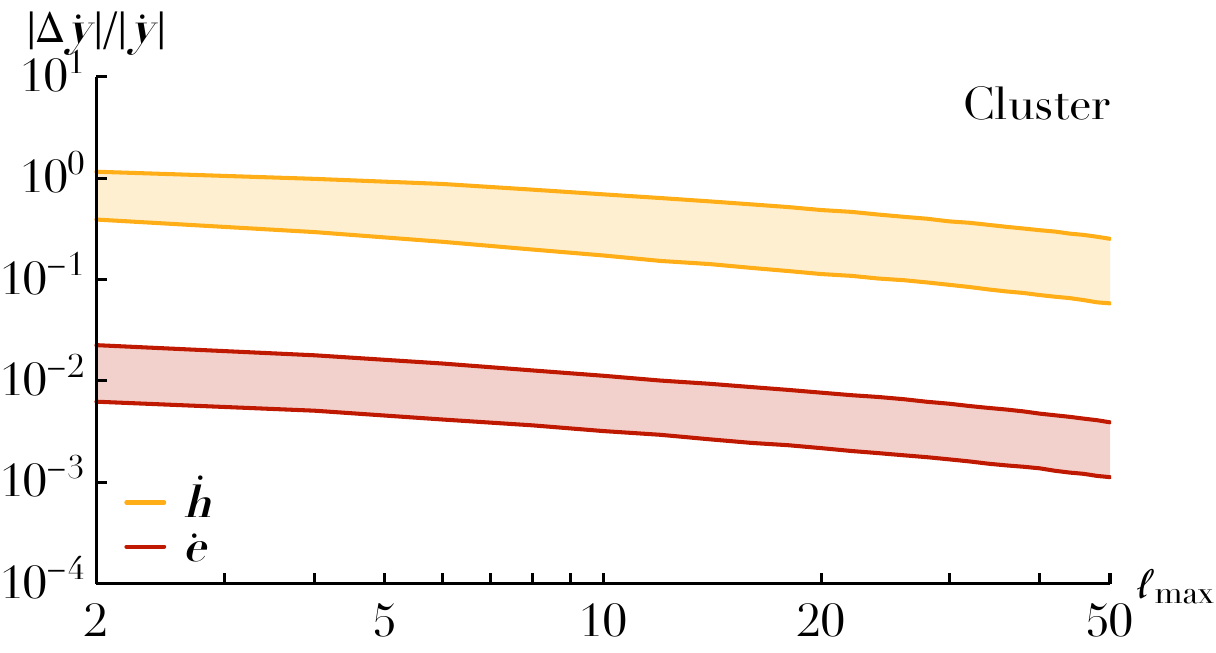}
\end{center}
\caption{Illustration of the relative error in $\dot{\vec{h}}$ and $\dot{\vec{e}}$ as a function of $\ellmax$, for a fixed number of nodes
($K=100$ for Kozai--Lidov and $K=20\,480$ for the cluster),
using the same conventions as in Figure~\ref{fig:errK}.
Relative errors are computed by comparison
to force calculations with $\ellmax=20$ for Kozai--Lidov,
and $\ellmax = 100$ for the cluster.
\label{fig:errlmax}}
\end{figure}

In Figure~\ref{fig:errlmax},
we investigate the relative errors in $\dot{\vec{h}}$
and $\dot{\vec{e}}$ 
as a function of $\ellmax$
for a fixed number of nodes $K$.
For the Kozai--Lidov simulations, we find that
the error decreases exponentially as $\re^{-1.6 \, \ellmax}$.
Such a rapid convergence stems from
the absence of any radial overlap
between the two stellar orbits.
Indeed, glancing back at equation~\eqref{Legendre_Expansion},
the radial dependence of a typical term of the Legendre expansion
is of the form $(r_{\min}/r_{\max})^{\ell} \le  \eta^{\ell}$,
with $\eta$ the largest distance ratio that occurs
as the inner and outer stars run through their orbits.
For the present setup,
one readily finds $\eta = 1/5$, see Appendix~\ref{sec:KozaiLidov}.
As a consequence, when truncated at order $\ellmax$,
the typical error is expected to scale
like $\eta^{\ellmax} \simeq \re^{-1.6 \, \ellmax}$,
matching the numerical measurement from Figure~\ref{fig:errlmax}.

In the same figure, we note that the convergence
w.r.t.\ $\ellmax$ is much slower for the $N=10^{4}$ clusters.
This is a consequence of the very slow convergence of the Legendre expansion~\eqref{Legendre_Expansion} in any regime where the wires overlap in radius.
We also note that $\dot{\vec{e}}$ is more accurate than $\dot{\vec{h}}$.
This is likely due to the analytical contribution from the Schwarzschild precession (see equation~\ref{1PN_Prec}),
which only affects $\dot{\vec{e}}$ and somewhat reduces the relative errors.
Let us finally emphasise that the truncation of the multipole approach effectively softens the interaction potential,
thereby avoiding the singularities that would otherwise
occur when wires cross.
As such, this method is well-suited to studying
scalar and vector resonant relaxation~\citep{RauchTremaine1996},
which are dominated by large-scale effects.

\begin{figure}
\begin{center}
\includegraphics[width=0.45\textwidth]{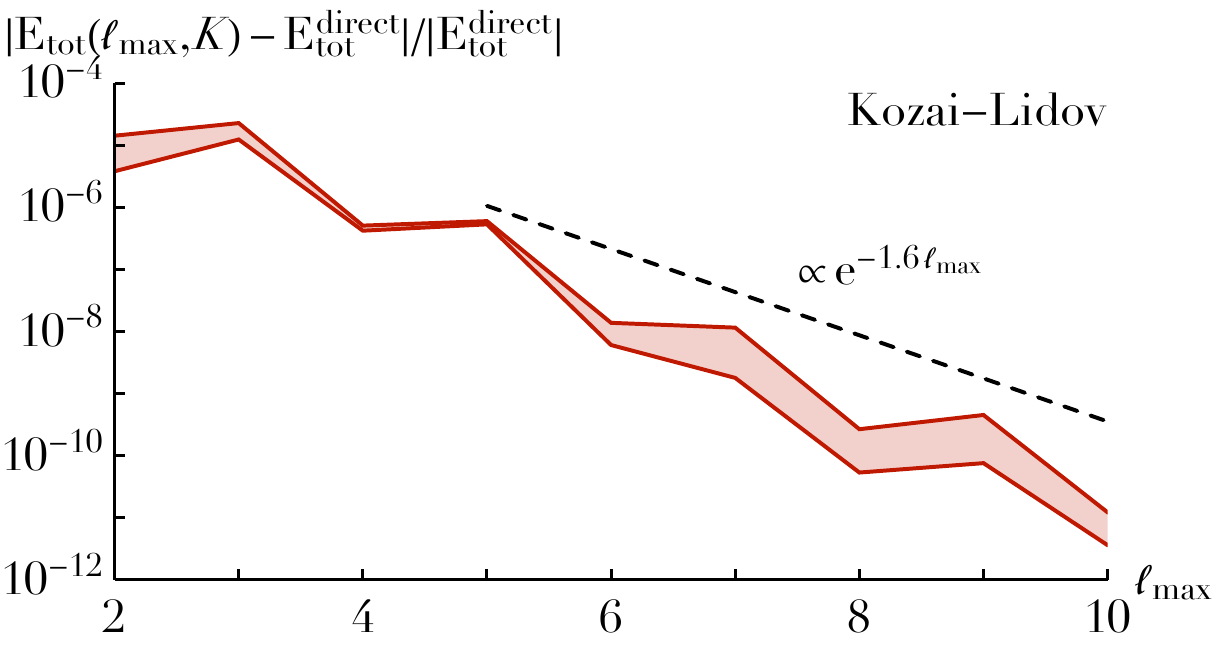}
\\
\includegraphics[width=0.45\textwidth]{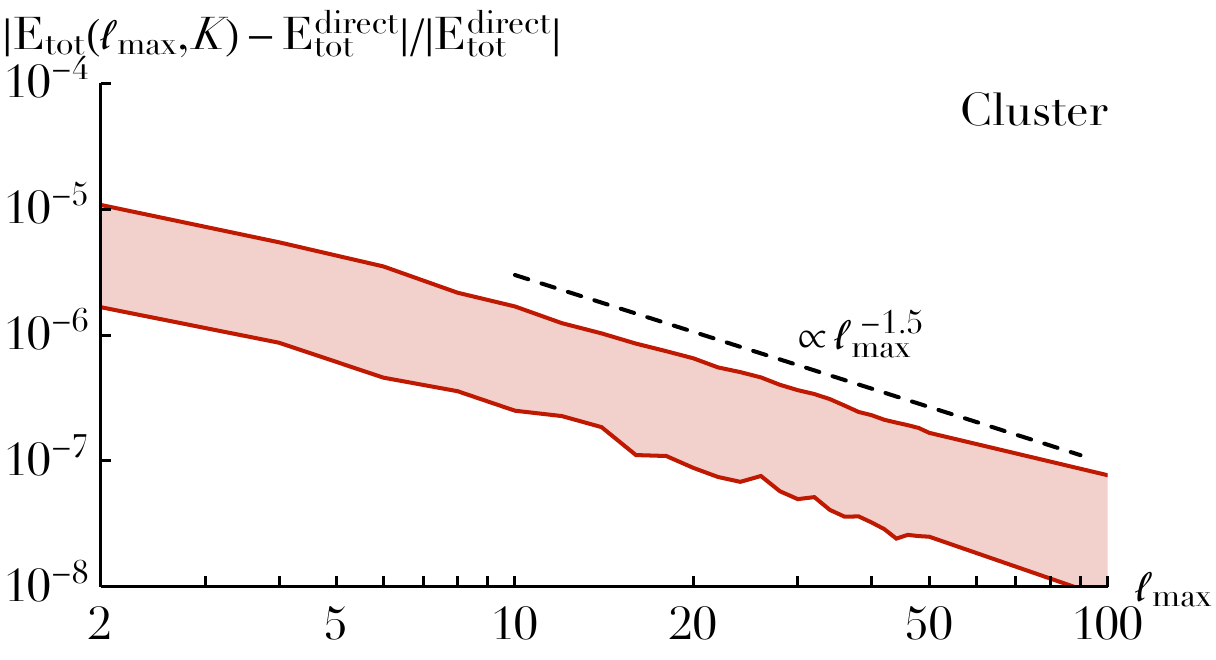}
\end{center}
\caption{Illustration of the relative error in the truncated total energy, $ \Etot (\ellmax,K)$,
as a function of the multipole truncation order $\ellmax$
for a fixed number of nodes ($K=100$ for Kozai--Lidov
and $K=20\,480$ for the cluster),
using the same convention as in Figure~\ref{fig:errEtot_K}.
The full total energy, $\Etotdirect$,
was computed by direct summation of equation~\eqref{Hrel,Hw}
using respectively $K=20\,480$ for the Kozai--Lidov simulations
and $K=1\,000$ for the clusters.
\label{fig:errEtot_lmax}}
\end{figure}

We investigate the same convergence in Figure~\ref{fig:errEtot_lmax}
by considering the dependence w.r.t.\ $\ellmax$
of the relative error between the truncated total energy, $\Etot (\ellmax,K)$,
and the full total energy, $\Etotdirect$.
For the Kozai--Lidov simulations,
we recover an exponential convergence w.r.t.\ $\ellmax$,
as $\re^{-1.6 \, \ellmax}$.
For the $N=10^4$ clusters,
where numerous orbits are radially overlapping,
we empirically find that $\Etot (\ellmax,K)$ converges
roughly like $\ellmax^{-1.5}$ w.r.t.\ the multipole truncation order.
This is compatible with the asymptotic scalings
predicted in Appendix~{B5} of~\cite{KocsisTremaine2015}
in the context of vector resonant relaxation.
Indeed, for radially overlapping non-coplanar orbits,
which mostly compose the present $N=10^{4}$ clusters,
a given harmonic, $\ell$, is found to give a contribution
to the total truncated energy scaling like $\ell^{-2.5}$.
Once summing over all the harmonics $0 \leq \ell \leq \ellmax$,
this ultimately leads to a relative error in $\Etot (\ellmax,K)$
scaling like $\ellmax^{-1.5}$.

\begin{figure}
\begin{center}
\includegraphics[width=0.45\textwidth]{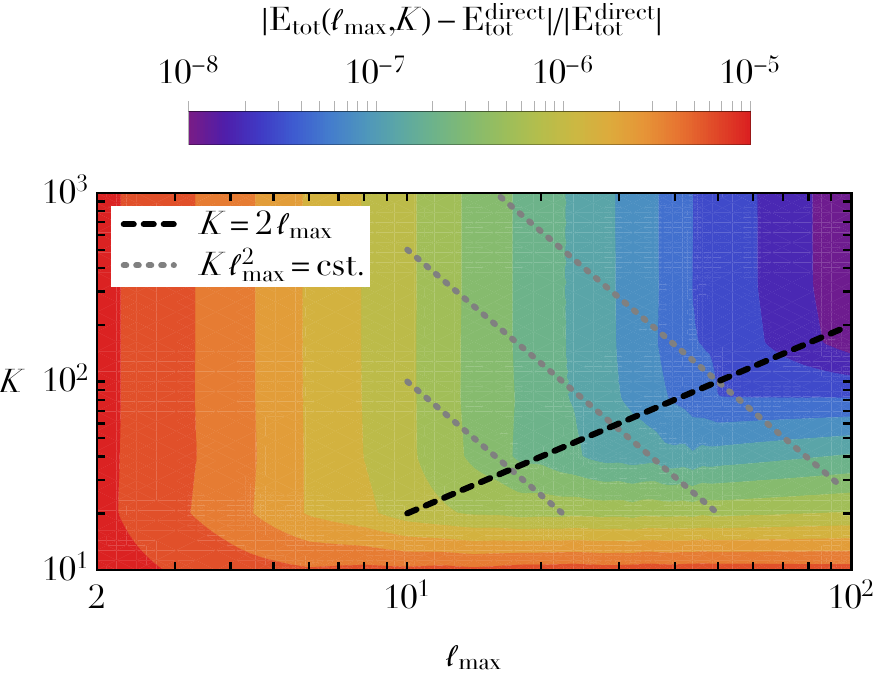}
\end{center}
\caption{Illustration of the median relative error
in the truncated total energy, $\Etot (\ellmax,K)$,
w.r.t.\ the full total energy, $\Etotdirect$,
as a function of $\ellmax$ and $K$
for the cluster simulations,
using the same parameters as in Figure~\ref{fig:errEtot_lmax}.
Errors are empirically found to saturate
for $K \simeq 2 \, \ellmax$ (black line),
while the computational complexity scales
like $O (K \ellmax^{2})$ (gray lines).
\label{fig:errEtot_lmaxK}}
\end{figure}

Finally, in Figure~\ref{fig:errEtot_lmaxK},
we illustrate the joint dependence
of the relative error
in the truncated total energy, $\Etot (\ellmax,K)$,
w.r.t.\ the full total energy, $\Etotdirect$,
as a function of $\ellmax$ and $K$
for the cluster simulations.
For the present system, we find that errors
saturate for $K \simeq 2 \, \ellmax$.
Because the computational complexity
of the discretised harmonic expansion
scales like $O (K \ellmax^{2})$,
for a fixed value of $\ellmax$,
ensuring converged estimations of $\Etot (\ellmax,K) $
requires $O (\ellmax^{3})$ operations.

\section{Time Integration}
\label{sec:IntegrationScheme}

In our application, the state of the system is described by the $N$ sets
$y = \{ \vec{b}_{+} , \vec{b}_{-} \}_{i}$. Given a timestep $\tau$, an integration scheme is a mapping $y_{n} \mapsto y_{n+1}$, where subscripts denote different time slices. Standard Runge--Kutta integration of the three-dimensional vectors $\vec{b}$ generally leads to violations of the constraints $|\vec{b}|=1$. Instead, we present here integration schemes that by construction exactly comply with these constraints. For convenience we define the velocity field $\vec{B}(y)=\dvec{h}\pm\dvec{e}$, so the equations of motions read $\dvec{b}=\vec{B}(y)$.

\subsection{Explicit Lie group methods}
\label{sec:explicit_scheme}

For a fixed value of the precession vector $\vec{\Omega}=\vec{b}\times\vec{B}$ (see equation~\ref{eq:Omega}) and an initial condition $\vec{b}_{0}$,
equation~\eqref{prec_b} can be integrated exactly for a duration $t$, via Rodrigues' rotation formula, to the new location
\begin{align}
    \label{Rodrigues}
    \vec{b} (t) & = \cos (\Omega t) \vec{b}_{0} + \sin (\Omega t) \uvec{\Omega} \times \vec{b}_{0} + [1 - \cos(\Omega t)] (\vec{b}_{0} \cdot \uvec{\Omega}) \uvec{\Omega}
\nonumber
\\
& = \phi \big[ t \vec{\Omega} \big] \circ \vec{b}_{0}
\end{align}
with $|\vec{b} (t)|=|\vec{b}_0|$.
In order to exactly preserve the constraints $|\vec{b}|=1$,
explicit integration schemes analogous to Runge--Kutta schemes can be devised via Lie group methods~\citep[see, e.g.\@, \S{IV.8} in][]{Hairer+2006}
by concatenating appropriately rotations using precession vectors
obtained at various intermediate stages.
In practice, we use \cite{MuntheKaas1999} integrators for their simplicity.
To do so, one rewrites the dynamics as
\begin{equation}
\vec{b} (t) = \phi \big[ \vec{U} (t) \big] \circ \vec{b}_{0} ,
\label{rewrite_MK}
\end{equation}
which imposes that $\vec{U}(t)$ evolves according to
\begin{equation}
\dvec{U} = \dphiinv_{\vec{U}} \circ \big\{ \vec{\Omega} \big( \phi \big[ \vec{U} \big] \circ \vec{b}_{0} \big) \big\} ,
\label{evol_U}
\end{equation}
with $\vec{\Omega} (\vec{b}) = \vec{b} \times \vec{B} (\vec{b})$
following equation~\eqref{eq:Omega}.
Here, $\dphiinv_{\vec{U}}$ is the inverse of the differential
of the rotation map~\citep[see Theorem~{3} in][]{MuntheKaas1999}.
It generically reads
\begin{equation}
\dphiinv_{\vec{U}} \circ \vec{\Omega} = \sum_{k = 0}^{+ \infty} \frac{B_{k}}{k!} \, \ad^{k}_{\vec{U}} \circ \vec{\Omega} ,
\label{exp_dphiinv}
\end{equation}
with $B_{k}$ the Bernoulli numbers and $\ad^{k}_{\vec{U}}$ the $k$-th power
of the adjoint operator.
In the present context, this operator simply becomes
\begin{equation}
\ad_{\vec{U}}^{0} \circ \vec{\Omega} = \vec{\Omega} ;
\quad
\ad_{\vec{U}}^{k} \circ \vec{\Omega} = \vec{U} \times \big( \ad_{\vec{U}}^{k-1} \circ \vec{\Omega} \big) .
\label{def_adj}
\end{equation}

A Munthe-Kaas scheme proceeds then by using a classical Runge-Kutta
scheme applied to $\vec{U}(t)$ and ``correcting''
the intermediate precession vectors with $\dphiinv_{\vec{U}}$
appropriately truncated~\citep[Theorem~{IV.8.5} in][]{Hairer+2006}.
For a second-order scheme,
one can use $\dphiinv_{\vec{U}} \circ \vec{\Omega} = \vec{\Omega}$,
from which one constructs the two-stage explicit midpoint rule (coined \texttt{MK2}). Starting from an initial state $\vec{b}_{n}$, it proceeds via
\begin{align}
    \label{def_MK2}
    \vec{\Omega}_{1} & = \vec{\Omega} (\vec{b}_{n}),
    \nonumber
    \\
    \vec{b}_{2} & = \phi \big[ \tfrac\tau2 \vec{\Omega}_{1} \big] \circ \vec{b}_{n} ,
    \nonumber
    \\
    \vec{\Omega}_{2} & = \vec{\Omega} (\vec{b}_{2}) ,
    \nonumber
    \\
    \vec{b}_{n+1} & = \phi \big[ \tau \vec{\Omega}_{2} \big] \circ \vec{b}_{n} .
\end{align}
Naturally the rotations are performed over all elements of $y$ simultaneously.

When constructing higher order methods, one must account for the fact that the rotations~\eqref{Rodrigues} do not commute,
i.e.\ better approximations of $\dphiinv_{\vec{U}}$ have to be used.
Here, we implement a four-stage scheme (coined \texttt{MK4})
based on the classical fourth-order Runge--Kutta scheme.
It reads
\begin{align}
    \label{def_MK4}
    \tvOmega_{1} & = \vec{\Omega} (\vec{b}_{n}),
    \nonumber
    \\
        \vec{b}_{2} & = \phi \big[ \tfrac\tau2 \tvOmega_{1} \big] \circ \vec{b}_{n} ,
    \nonumber
    \\
    \tvOmega_{2} & = \dphiinv_{\tfrac\tau2 \tvOmega_{1}} \circ \vec{\Omega} (\vec{b}_{2}) ,
    \nonumber
    \\
       \vec{b}_{3} & = \phi \big[ \tfrac\tau2 \tvOmega_{2} \big] \circ \vec{b}_{n} ,
    \nonumber
    \\
        \tvOmega_{3} & = \dphiinv_{\tfrac\tau2 \tvOmega_{2}} \circ \vec{\Omega} (\vec{b}_{3}) ,
    \nonumber
    \\
      \vec{b}_{4} & = \phi \big[ \tau \tvOmega_{3} \big] \circ \vec{b}_{n} ,
    \nonumber
    \\
       \tvOmega_{4} & = \dphiinv_{ \tau \tvOmega_{3}} \circ \vec{\Omega} (\vec{b}_{4}) ,
    \nonumber
    \\
    \vec{b}_{n+1} & = \phi \big[ \tau \big( \tfrac16 \tvOmega_{1} + \tfrac13 \tvOmega_{2} + \tfrac13 \tvOmega_{3} + \tfrac16 \tvOmega_{4} \big) \big] \circ \vec{b}_{n} ,
\end{align}
where $\dphiinv_{\vec{U}}$
is truncated at second order in $\vec{U}$
so that equation~\eqref{exp_dphiinv} becomes
\begin{equation}
\dphiinv_{\vec{U}} \circ \vec{\Omega} = \vec{\Omega} - \tfrac12 \vec{U} \times \vec{\Omega} + \tfrac{1}{12} \vec{U} \times \big( \vec{U} \times \vec{\Omega} \big) .
\label{def_dexpinv}
\end{equation}
The schemes~\eqref{def_MK2} and~\eqref{def_MK4} are (i) explicit, (ii) conserve $|\vec{b}|=1$ exactly\footnote{We systematically perform the re-normalisation $\vec{b} \leftarrow \vec{b}/|\vec{b}|$ after every evaluation of equation~\eqref{Rodrigues} to prevent a drift of $|\vec{b}|$ from round-off errors.}, (iii) require two or four computations of the derivatives, respectively, and (iv) are, respectively, second- and fourth-order accurate.
Because they are direct translations of usual Runge--Kutta methods,
it is straightforward
to design Munthe-Kaas schemes
of higher order.

\subsection{Symplectic scheme}
\label{sec:symplectic_scheme}
Following~\cite{McLachlan+2014}, we also consider an integrator relying on the spherical midpoint method (coined \texttt{MD2}). This is based on the implicit relation
\begin{equation}
    \frac{\vec{b}_{n+1} - \vec{b}_{n}}{\tau} = \vec{B} \left( \frac{\vec{b}_{n+1} + \vec{b}_{n}}{| \vec{b}_{n+1} + \vec{b}_{n} |} \right) .
    \label{def_MD2}
\end{equation}
This relation can be solved via the fixed-point iteration
\begin{equation}
    \label{eq:fixed_point}
    \vec{b}_{n+1}^{(0)} = \vec{b}_{n} ;
    \qquad
    \vec{b}_{n+1}^{(k+1)} = \vec{b}_{n} + \tau \, \vec{B} \left( \frac{\vec{b}_{n} + \vec{b}_{n+1}^{(k)}}{| \vec{b}_{n} + \vec{b}_{n+1}^{(k)} |} \right) ,
\end{equation}
terminating when $|| \vec{b}_{n+1}^{(k+1)} - \vec{b}_{n+1}^{(k)} ||_{\infty} \leq \epsilon$, for a given tolerance $\epsilon$. Naturally, these iterations are performed for all the elements of $y$ simultaneously. In practice, we impose $\epsilon = 2.2 \times 10^{-15}$, which typically requires 5--12 iterations.

The scheme from equation~\eqref{def_MD2} is (i) second-order accurate; (ii) implicit; (iii) exactly conserves the constraints from equation~\eqref{constraints_bp_bm}\footnote{The outcome of the fixed-point search in equation~\eqref{eq:fixed_point} is systematically renormalised via $\vec{b} \leftarrow \vec{b}/|\vec{b}|$ to prevent a drift of $|\vec{b}|$ from round-off errors.}; (iv) symplectic~\citep{McLachlan+2014}; and (iv) conserves $\Ltot = \sum_{i} \Lambda_{i} \vec{h}_{i}$, as it is a global linear invariant.
Because \texttt{MD2} is a symmetric scheme, one can
use symmetric composition techniques~\citep[see, e.g.\@, \S{V.3.2} in][]{Hairer+2006} to devise higher order symplectic integrators.

The \texttt{MK2}, \texttt{MK4} and \texttt{MD2} schemes
conserve the constraints for the vectors $\vec{h},\,\vec{e}$ to machine precision, which as far as we know none of the previous studies~\citep[e.g.\@,][]{ToumaTremaine2009, TremaineTouma2009, Hamers2016} have managed.
 
\subsection{Convergence of the time integrations}
\label{sec:ConvergenceTime}
\begin{figure}
    \begin{center}
    \includegraphics[width=0.45\textwidth]{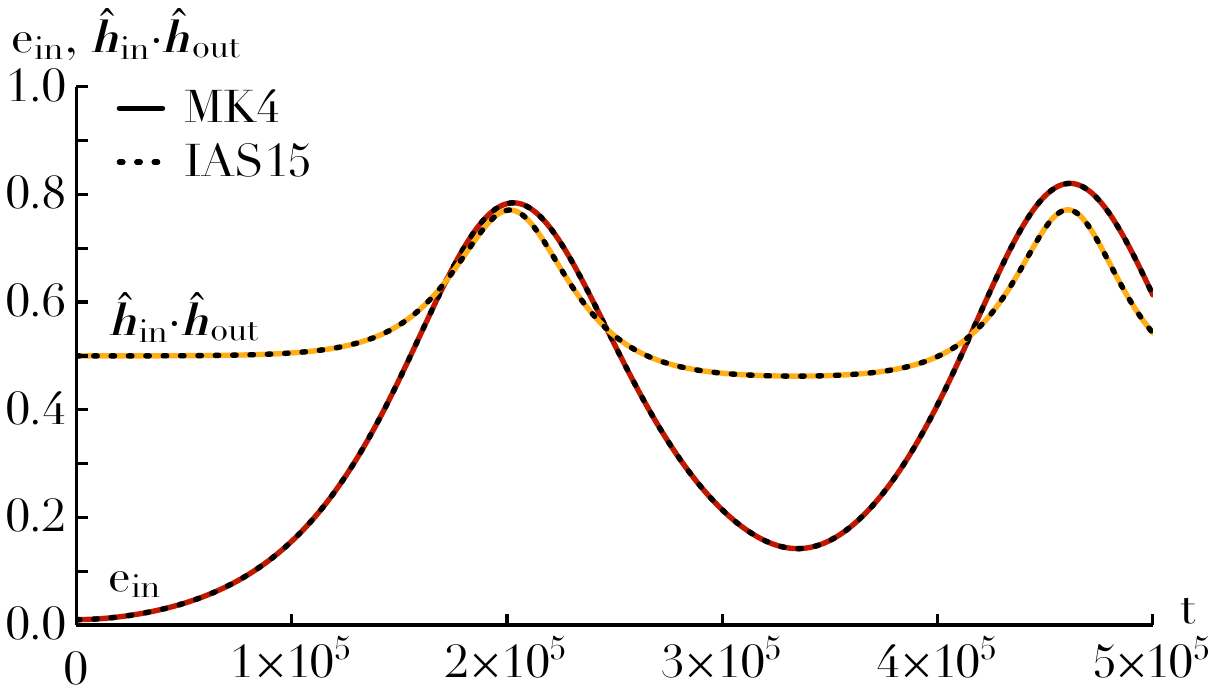}
    \end{center}
\caption{Illustration in the Kozai--Lidov simulations
    of the evolution of the inner star's eccentricity, $e_{\mathrm{in}}$, as well as its inclination w.r.t.\ the outer star, $\uvec{h}_{\mathrm{in}} \cdot \uvec{h}_{\mathrm{out}}$. Full lines correspond to the multipole integration (with the \texttt{MK4} scheme from equation~\ref{def_MK4}), while dashed lines were obtained by direct integration using the \texttt{IAS15} integrator~\citep{Rein+2015}. See Appendix~\ref{sec:KozaiLidov} for details.
    \label{fig:rebound}}
\end{figure}

As a first check of the sanity of the present algorithm, we compare in Figure~\ref{fig:rebound} the time-evolution of the Kozai--Lidov system as predicted by the multipole approach and a direct integration of the associated 3-body problem.
Both methods predict similar oscillations of the inner star\footnote{When represented in Figure~\ref{fig:rebound},
the \texttt{MK2}, \texttt{MK4} and \texttt{MD2} integrations are indistinguishable.}. On average, the direct integration used an integration timestep of the order $\tau \simeq 10^{-4}$, while for the multipole integration, benefiting from its explicit orbit average, we used $\tau = 10$, i.e.\ a significantly larger timestep.

\begin{figure}
    \begin{center}
        \includegraphics[width=0.45\textwidth]{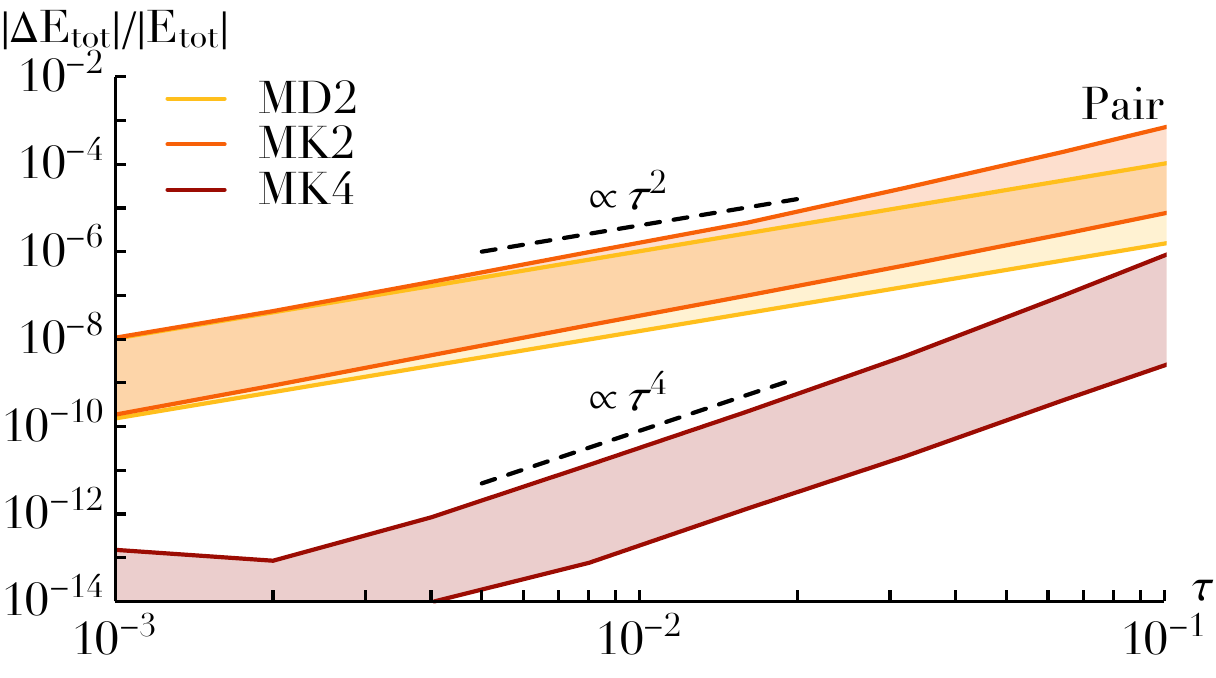}
    \end{center}
    \caption{Illustration of the relative error in the total energy $\Etot$ w.r.t.\ its initial value after a fixed time $T = 1.28$, as a function of the considered integration timestep, $\tau$. The colored regions correspond to the 16\% and 84\% levels over 500 independent realisations of the ``Pair'' Hamiltonian. We compute the relative errors by comparison with the initial total energy. See Appendix~\ref{sec:Pair} for details.
    \label{fig:errPairs_tau}}
\end{figure}

In order to further assess the performance of the integration schemes, we now consider the simple two-body ``Pair'' Hamiltonian from Appendix~\ref{sec:Pair}. In Figure~\ref{fig:errPairs_tau}, we illustrate the dependence of the relative error
in $\Etot$ after a finite time, as a function of the integration timestep $\tau$.
We recover that \texttt{MK2} and \texttt{MD2} are second-order accurate with a finite-time error converging like $O(\tau^{2})$,
while \texttt{MK4} is fourth-order accurate.

\begin{figure}
    \begin{center}
        \includegraphics[width=0.45\textwidth]{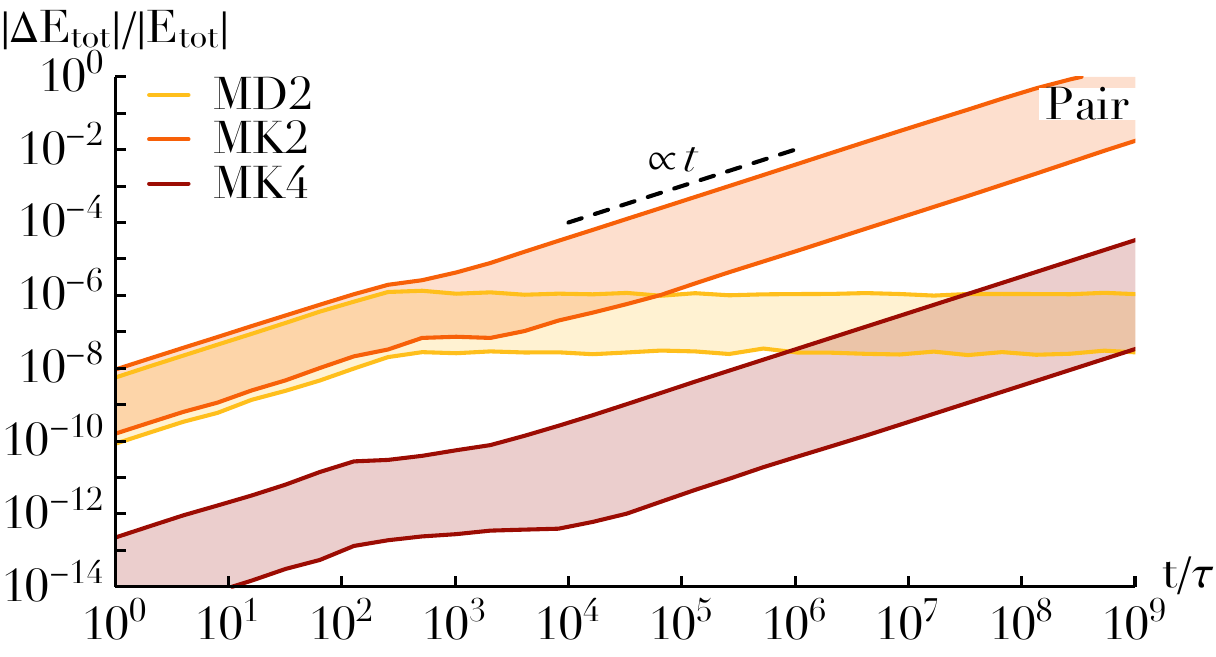}
        \\
        \includegraphics[width=0.45\textwidth]{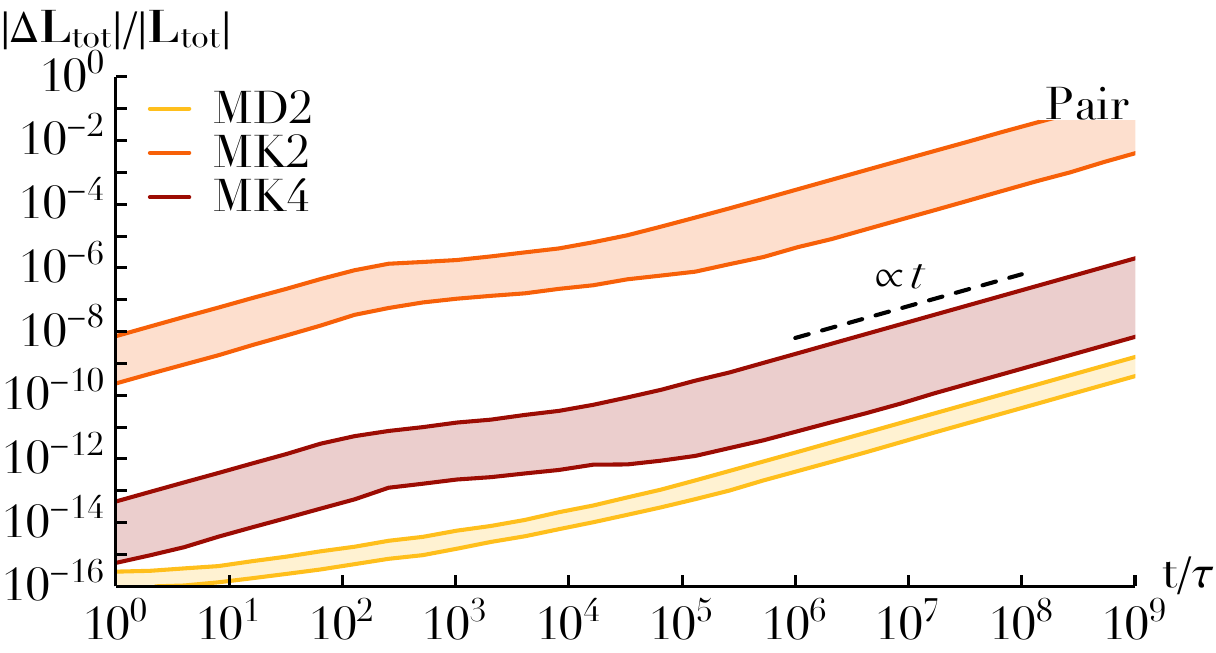}
    \end{center}
    \caption{Illustration of the relative error in the total energy $\Etot$ (top) and total angular momentum $\Ltot$ (bottom) w.r.t.\ their initial values, as a function of the number of integration steps, $t/\tau$, using the same convention as in Figure~\ref{fig:errPairs_tau}. Here, we used $\tau = 10^{-2}$ as our integration timestep.
    \label{fig:errPairs_time}}
\end{figure}

\begin{figure}
    \begin{center}
        \includegraphics[width=0.45\textwidth]{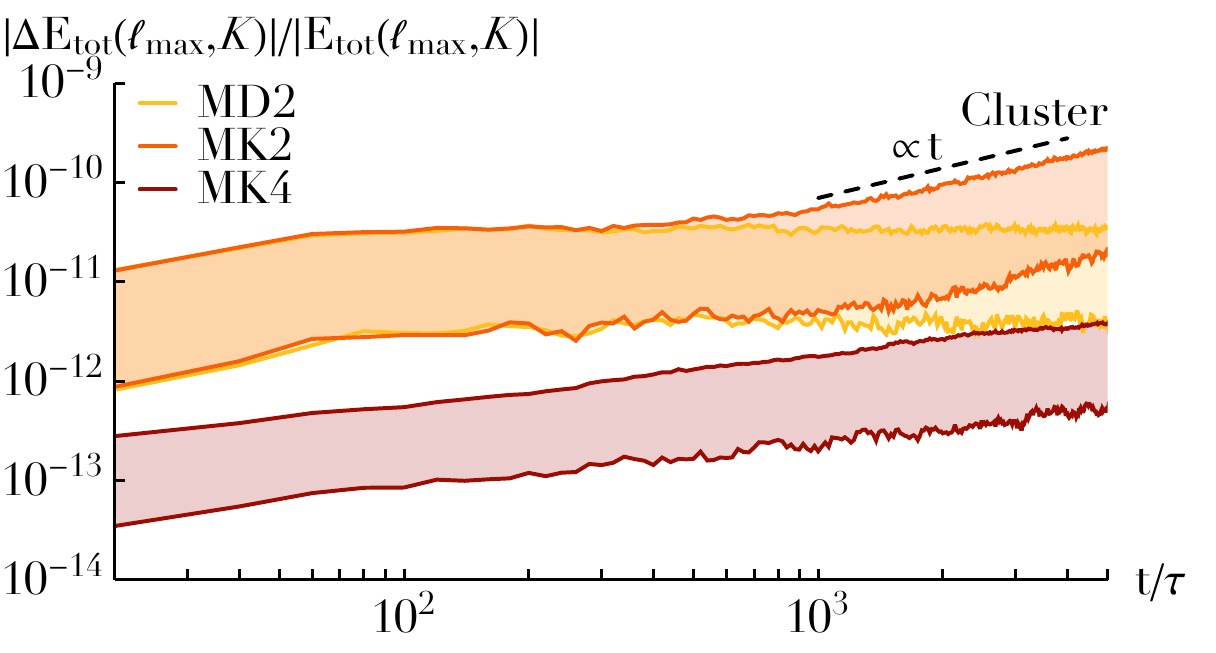}
        \\
        \includegraphics[width=0.45\textwidth]{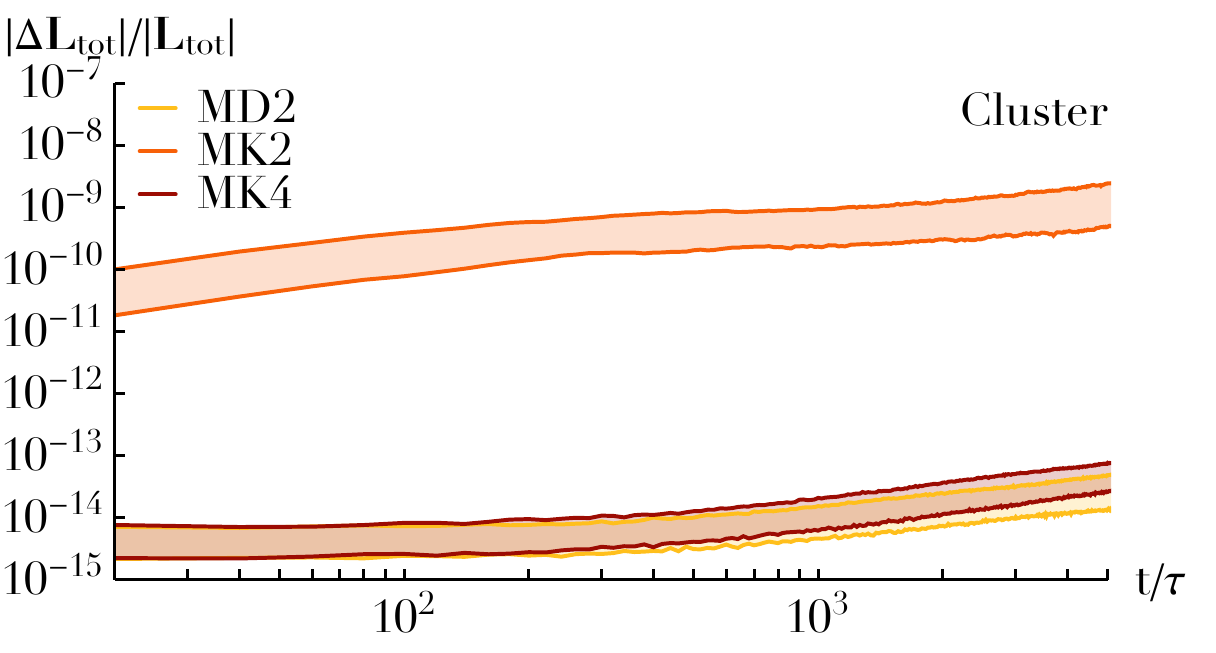}
    \end{center}
    \caption{Illustration of the relative error in the truncated total energy $\Etot (\ellmax,K)$ (top) and total angular momentum $\Ltot$ w.r.t.\ their initial values, as a function of the number of integration steps, $t/\tau$, for the $N=10^{4}$ clusters, using the same convention as in Figure~\ref{fig:errPairs_time}. Here, we used $\ellmax = 10$, $K = 100$,
    and $\tau = 10^{-2} \, \taudyn$ as our integration timestep, see Appendix~\ref{sec:PhysicalModel}.
    \label{fig:errEtotLtot_time}}
\end{figure}

In Figure~\ref{fig:errPairs_time}, for the same Hamiltonian, we investigate the long-time trends for the errors in $\Etot$ and $\Ltot$.
As expected, the explicit schemes show errors growing like $t$. There are two main improvements in the symplectic scheme: (i) the error in $\Etot$ is bounded on long timescales; (ii) the error in $\Ltot$ only grows via the accumulation of round-off errors.
Similarly, in Figure~\ref{fig:errEtotLtot_time}, we investigate the errors in $\Etot(\ellmax,K)$ and $\Ltot$ for the $N=10^4$ clusters.
This figure exhibits similar asymptotic trends as in Figure~\ref{fig:errPairs_time}.

\begin{figure}
    \begin{center}
    \includegraphics[width=0.45\textwidth]{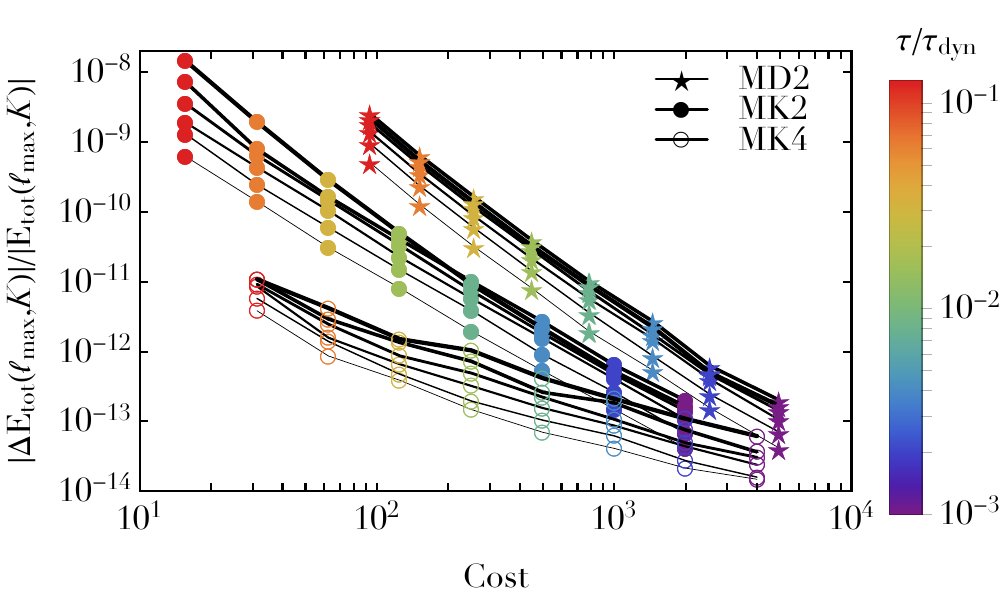}
    \\
    \includegraphics[width=0.45\textwidth]{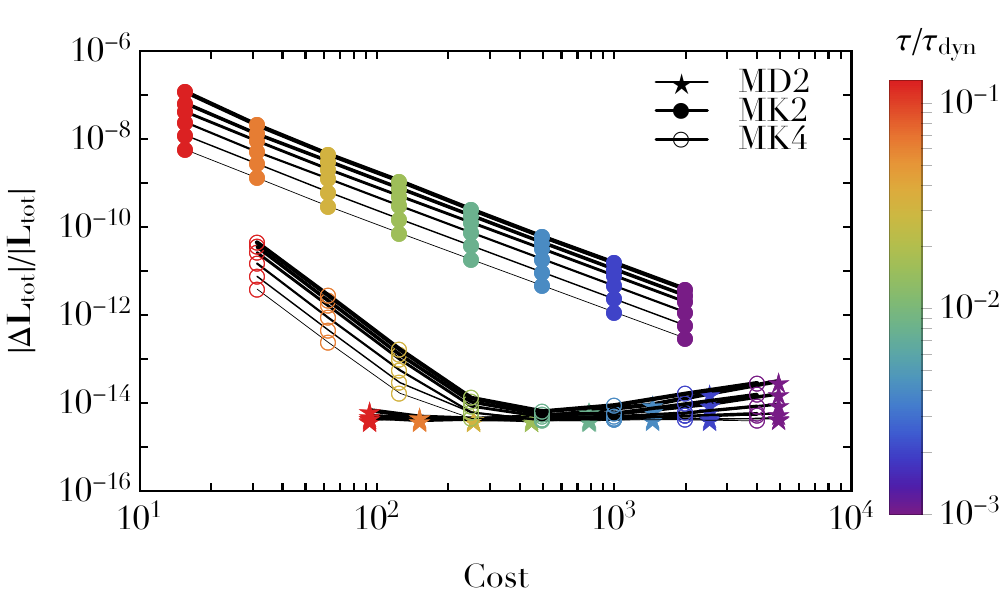}
    \end{center}
    \caption{Error-cost relation for the cluster simulations. Simulations were performed with (i) the symplectic scheme (\texttt{MD2}; star symbols)
    or  the explicit ones (\texttt{MK2} and \texttt{MK4}; circle symbols); with (ii) different integration timesteps, $\tau/\taudyn = 2^{k} \times 10^{-3}$ ($0 \leq k \leq 7$; different colors); (iii) up to different finite times, $T/\taudyn = 0.128 \times 2^{-k}$ ($0 \leq k \leq 5$): the larger $T$, the thicker the line. The cost of a given simulation is estimated via $\avforces/(\tau/\taudyn)$, with $\avforces$ the average number of force evaluations per integration timestep, e.g.\@, two for \texttt{MK2}. Median errors in $\Etot (\ellmax,K)$ (top) and $\Ltot$ (bottom) are estimated over $128$ realisations, with $\ellmax = 10$ and $K=100$.
\label{fig:Efficiency}}
\end{figure}

Finally, in Figure~\ref{fig:Efficiency},
we illustrate the error-cost relation
in the cluster simulations.
The cost of a given simulation is estimated via $\avforces/(\tau/\taudyn)$, with $\avforces$ the average number of force evaluations per integration timestep
and $\taudyn$ the shortest dynamical time of the clusters at hand (see equation~\ref{choice_tau}). While the explicit schemes always perform a constant number of force evaluations per timestep, the symplectic scheme typically requires from 5 evaluations (smallest $\tau$) up to $12$ (largest $\tau$).
As one reduces $\tau$,
the error in $\Etot (\ellmax,K)$
keeps shrinking,
illustrating that our discretisation scheme
indeed corresponds to a Hamiltonian system
(equation~\ref{eq:cons_Etot}).
Finally, as already hinted in Figure~\ref{fig:errEtotLtot_time}, for the present rather short-duration integrations, at a given cost, the explicit schemes outperforms the symplectic one regarding the conservation of $\Etot(\ellmax,K)$, while the converse holds regarding $\Ltot$.

\section{Scalar Resonant Relaxation}
\label{sec:RR}

Using the method described above, we may now use our simulations of a $N=10^{4}$ cluster to model the dynamics around the supermassive \BH\ SgrA*. We use these simulations to study scalar resonant relaxation~\citep{RauchTremaine1996}, the process through which Keplerian wires can relax their eccentricities through secular interactions with one another. We measure the associated diffusion coefficients, and compare them with the theoretical predictions put forward by~\cite{BarOrFouvry2018}. All the details for these measurements and predictions are spelled out in Appendix~\ref{sec:PhysicalModel}.

In Figure~\ref{fig:SRR}, we present the main result of this comparison, the (finite-time) diffusion coefficients $ D_{hh} = \{ [\Delta h(T)]^{2} \} / T$, where $\{ \,\cdot\, \}$ stands for an ensemble average, and we used $T \!=\! 25 \, \mathrm{kyr}$, $\ellmax \!=\! 10$, and $K \!=\! 100$.
\begin{figure*}
\begin{center}
\includegraphics[width=0.85\textwidth]{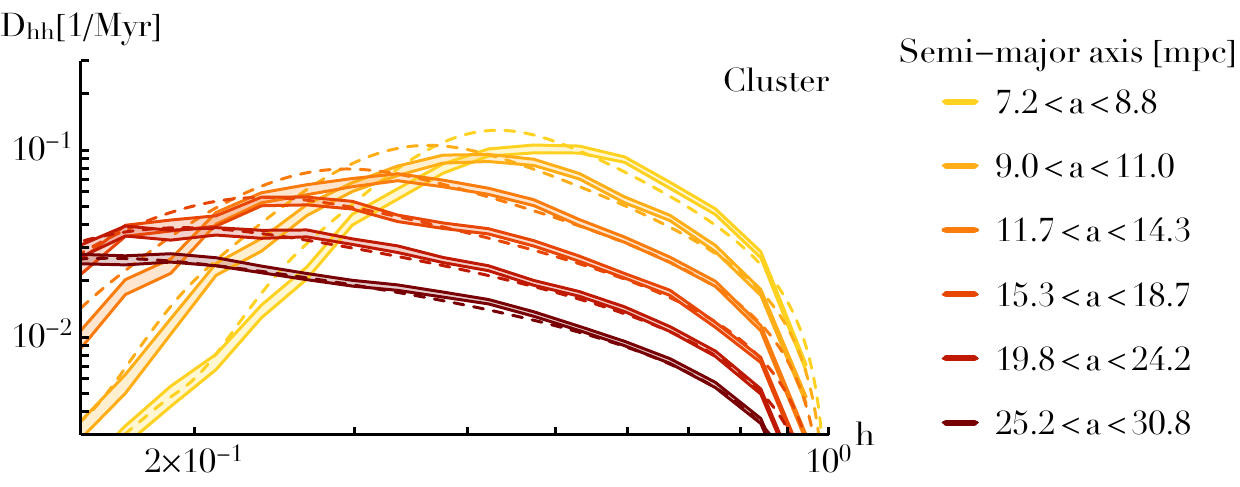}
\end{center}
\caption{Illustration of the (finite-time) diffusion coefficients $D_{hh}=\{ [\Delta h(T)]^{2} \} /T$, with $h=\sqrt{1 - e^{2}}$ in a galactic nucleus mimicking SgrA*. Wires are split in different bins of semi-major axes ($a$) in milliparsecs, represented with different colours. The dashed curves correspond to the theoretical predictions following the formalism of~\cite{BarOrFouvry2018}. The thick curves correspond to the numerical measurements performed in the multipole simulations and their associated bootstrap errors. See Appendix~\ref{sec:PhysicalModel} for details.
\label{fig:SRR}}
\end{figure*}
This figure shows a good agreement between the diffusion coefficients inferred from the numerical simulations and from kinetic theory. It is likely that the remaining mismatch stems from the difficulty of measuring/predicting secular diffusion coefficients on such a short finite time, $T$, see equation~{(24)} in~\cite{BarOrFouvry2018}. Figure~\ref{fig:SRR} further illustrates the sanity of the present algorithm, which reproduced the intricate resonant dynamical interactions of Keplerian wires in a galactic nucleus. Of course, the present application is only a first illustration of the physical mechanisms that can be investigated with our multipole method. Astrophysical applications to more diverse physical systems will be the subject of future work.

\section{Discussion}
\label{sec:Discussion}

First, we note that the computation of the rates of change $( \dot{\vec{h}}, \dot{\vec{e}} )$ for all the nodes has a computational complexity in $O (N K \ellmax^{2})$. This is the main benefit of the present method which, at the cost of (truncated) harmonic expansions and discretised orbit averages, has a computational complexity scaling linearly in the total number of wires.

We also note that the computations of $\{ \dot{\vec{h}} , \dot{\vec{e}} \}_{\all}$ and ${ \{ \dot{\vec{h}} , \dot{\vec{e}} \}_{\self} }$ are completely independent (and of similar computational difficulty), so that they can be easily be performed in parallel. Similarly, for each of these computations, the recurrences over $P_\ell^m$ and $Q_\ell^m$, i.e.\ steps 4 and 6 of the algorithm given in Table~\ref{tab:algorithm}, are also independent one from another (and of similar computational difficulty), so that they can also be performed in parallel. As a consequence, one can easily benefit from a parallelisation of the previous algorithm over four cores, to compute in parallel $\{ \dot{\vec{h}}, \dot{\vec{e}} \}_{\all}$ and $\{ \dot{\vec{h}}, \dot{\vec{e}} \}_{\self}$ due to nodes at smaller as well as larger radii.

Further parallelisation requires the parallel computation of the prefix sums from equations~\eqref{eqs:P,Q} and is much more complex, but not impossible. We plan to investigate the application of the parallel scan algorithm by \cite{LadnerFischer1980} in a future study. 

Since the spherical harmonic expansion is isotropic, each wire's self-gravity, $\vec{f}^{\mathrm{self}}$, produces no torques and hence only affects each wire's apsidal precession rates, $\dot{\vec{e}}\cdot\uvec{k}/e$. In addition, these terms scale trivially with semi-major axis and hence depend non-trivially only on eccentricity. An alternative possibility for correcting for these self-gravity contributions is therefore to use the scaling in $a$ and a one-dimensional interpolation in $e$ from a table computed in advance for the self-gravity-induced apsidal precession at a given $a$. In practice, this could essentially halve the computational costs. Since our present implementation shares these costs between 4 CPUs, such a scheme would reduce this to 2 CPUs but not speed-up the computation overall. However, in an implementation featuring the parallel scan (see above) this would reduce the wall-clock time at fixed number of CPUs. 

A limitation of the present approach is the difficulty of using different timesteps for different wires. Indeed, in order to compute the prefix sums~\eqref{eqs:P,Q}, it is essential to scroll through the entire population of nodes. Similarly, it would be of interest to investigate adapting the number of nodes $K$ as a function of the wires' orbital parameters, in particular eccentricity.
These aspects deserve further analyses.

Our algorithm has problems with nodes at identical or very similar radii. Indeed, the spherical harmonic expansion~\eqref{Legendre_Expansion} fails to converge in the limit $r_i\to r_{\!j}$, even if $|\vec{r}_i-\vec{r}_{\!j}|>0$, i.e., even if the nodes are spatially well separated. This problem can be solved by reverting to direct summation of the forces between nodes with near-identical radii: this only requires a slight modification of our algorithm. Such an approach also calls for force softening in order to avoid artificially large forces between nodes that are much closer to one another than to their respective wire-companions. This observation was already made in Figure~{4} of~\cite{Dehnen2014}. In essence, our present setup has a limiting opening angle $\theta=1$, where the maximum errors hardly depend on $\ellmax$. For the errors to decrease significantly and hence offer a better efficiency at a given accuracy, one needs $\theta<1$, which can be achieved by using direct summation over nodes with similar radii. We reserve the development and testing of this more general approach for a future publication. In the present work, we merely avoid situations with nodes at identical radii (except for nodes on the same wire, where it does not matter) by not having two wires with exactly the same $(a,e)$.

The present multipole method differs from an alternative approach based on Gau{\ss}' method~\citep{ToumaTremaine2009}, which does not rely on a harmonic expansion. As a result, Gau{\ss}' method must unavoidably compute all the pairwise interactions between the wires, i.e., it has a computational complexity that scales quadratically in the number of wires. More precisely, its computational complexity scales like $O(N^{2}\KG)$, with $\KG$ the typical number of sampling points used to perform the second orbit average, which cannot be performed analytically. Gau{\ss}' method has several  strengths: (i) it is very well suited to perform time integration of orbit-averaged problems with a small number of wires; (ii) because the computation of the orbit-averaged force between two wires is a numerically expensive calculation, i.e., it has a computational complexity in $O(\KG)$, the method can also significantly benefit from parallelisation; (iii) because it relies on the computation of pairwise interactions, Gau{\ss}' method can also be further parallelised over pairs of stars, contrary to the present multipole method; (iv) Gau{\ss}' method is straightforward to apply in the case of softened interactions~\citep{ToumaTremaine2009}. One of the main drawbacks of Gau{\ss}' method is that its computational difficulty gets prohibitively large as $N$ gets larger, an aspect that the multipole method alleviates by design.

In Figure~\ref{fig:errPairs_time}, we noted that the accumulation of round-off errors in the symplectic scheme led to a biased growth of the error in $\Ltot$,
scaling like $t$. It would be worthwhile to improve the algorithm to always ensure an unbiased error growth in $\sqrt{t}$, following~\citeauthor{Brouwer1937}'s (\citeyear{Brouwer1937}) law, while still complying (exactly) with the constraints from equation~\eqref{constraints_bp_bm}. Similarly, one could investigate further the design of higher order structure-preserving integration schemes~\citep[see, e.g.\@,][]{Hairer+2006}.

\section{Conclusion}
\label{sec:Conclusion}

In this paper, we described the use of a multipole expansion to perform orbit-averaged simulations of stellar dynamics in galactic nuclei. We also presented integration schemes complying exactly with the system's orbital constraints. We used this method to recover the predicted statistical properties of resonant relaxation of stellar eccentricities. This algorithm has two main advantages, namely relying on a explicit orbit average over the fast Keplerian motion to reduce the range of dynamical times in the system, and offering, through the multipole expansion, a computational complexity scaling linearly with the total number of particles. The code will be made available on request.

The present work is only a first step towards the development of faster integration methods for the dynamics of galactic nuclei, here seen as an archetype of dynamically degenerate systems, i.e., systems exhibiting a global resonance condition in their orbital frequencies. As highlighted in the main text, there are three main limitations to the present method. First, even after the orbit average over the fast Keplerian motion induced by the central \BH\@, the range of dynamical times in a typical nucleus still remains very significant. This is partly a consequence of the divergence of the relativistic in-plane precession frequency for very eccentric wires. Similarly, the system's dynamical range is also enlarged by the ranges in semi-major axes and individual masses. For these reasons, even while using the present method, long time integrations still remain challenging. A second limitation of the present method is associated with its difficult parallelisation, in particular the prefix sums, as well as with the difficulty of tailoring it to multi-timestep schemes.
A third limitation stems from the current lack of any force softening and direct summations between nearby nodes. Such improvements are expected to increase the numerical accuracy and stability of the scheme.
All these aspects deserve further investigation.

Finally, when augmented by a second orbit average over the in-plane precession, the present methods can also be used to perform fast simulations of the dynamics of massive \emph{annuli} in galactic nuclei, which undergo vector resonant relaxation~\citep{KocsisTremaine2015,SzolgyenKocsis2018}.

\acknowledgments
We thank the anonymous referee for a very detailed and insightful report.
JBF acknowledges support from Program number HST-HF2-51374
which was provided by NASA through a grant from the Space Telescope Science Institute, which is operated by the Association of Universities for Research in Astronomy, Incorporated, under NASA contract NAS5--26555. ST acknowledges the support of the Natural Sciences and Engineering Research Council of Canada (NSERC), funding reference number RGPIN-2020-03885. BB is supported by the Martin A. and Helen Chooljian Membership at the Institute for Advanced Study. This work is partially supported by grant Segal ANR-19-CE31-0017 of the French Agence Nationale de la Recherche, and by the Idex Sorbonne Universit\'e. This work has made use of the Infinity Cluster hosted by the Institut d'Astrophysique de Paris. We thank St\'ephane Rouberol for running this cluster smoothly.

\appendix

\section{Derivation of the discretised equations of motion}
\label{sec:EOM:disc}
Here, we detail the derivation of equations~\eqref{EOM_mu} and~\eqref{EOM_r} from Milankovitch's relations~\eqref{eq:Milankovitch}  and the discretised Hamiltonian $\av{}{H_\star}$ as given in equation~\eqref{Htot}.

We first consider the terms induced by the dependencies of $\av{}{H_\star}$ on $\vec{h}_i$ and $\vec{e}_i$ through the node masses $\mu_{ik}$. From equation~\eqref{node:mass},
\begin{align}
    \frac{\partial\mu_{ik}}{\partial\vec{h}_{i}}
    &=0,
    &
    \frac{\partial\mu_{ik}}{\partial\vec{e}_{i}} &= -\bar{\mu}_i \cos E_k\, \uvec{e}_{i}.
\end{align}
with $\bar{\mu}_i\equiv m_i/K$, and therefore from equation~\eqref{Htot}
\begin{align}
    \frac{\partial\av{}{H_\star}}{\partial\vec{h}_i}\bigg|_{\mu} &=0,
    &
    \frac{\partial\av{}{H_\star}}{\partial\vec{e}_i}\bigg|_{\mu} &= -\uvec{e}_i \sum_k \bar{\mu}_i \,\psi_{ik}\,\cos E_k.
\end{align}
Milankovitch's equations~\eqref{eq:Milankovitch} allow us then to obtain equations~\eqref{EOM_mu}.

Next, we derive the terms induced by the dependencies of $\av{}{H_\star}$ on $\vec{h}_i$ and $\vec{e}_i$ through the node positions $\vec{r}_{ik}$. For a fixed vector $\vec{z}$, equation~\eqref{exp_br} gives us the derivative of the node positions via\footnote{Since $e = |\vec{e}| = \sqrt{1-\vec{h}^{2}}$, the eccentricity dependence can equivalently be considered a dependence on $\vec{h}$. Indeed, Milankovitch's equations~\eqref{eq:Milankovitch} benefit from a gauge freedom in regard to the constraints from equation~\eqref{constraints_bj_be} \citep[see the discussion after equation~\eqref{eq:Milankovitch:b} or Appendix~{A} in][]{TremaineTouma2009}, so that the final expressions for $\dot{\vec{h}}$ and $\dot{\vec{e}}$ are unaffected by this choice.}
\begin{align}
\label{eq:dr_dhde}
\frac{\partial (\vec{z} \cdot \vec{r}_{ik})}{\partial \vec{h}_{i}} & = a_{i} \sin E_k \, \uvec{e}_{i} \times \vec{z} ,
&
\frac{\partial (\vec{z} \cdot \vec{r}_{ik})}{\partial \vec{e}_{i}} & = - a_{i} \vec{z} + \frac{a_{i}}{e_{i}} \cos E_{k} \, (\vec{z} - \vec{z} \cdot \uvec{e}_{i} \, \uvec{e}_{i}) - \frac{a_{i}}{e_{i}} \sin E_{k} \, [ (\vec{h}_{i} \times \vec{z}) - (\vec{h}_{i} \times \vec{z}) \cdot \uvec{e}_{i} \uvec{e}_{i} ] .
\end{align}
Differentiating equation~\eqref{Htot} and inserting these relations, we get
\begin{align}
\label{eq:dHtot_dhde}
\frac{\partial \av{}{H_{\star}}}{\partial \vec{h}_{i}} \bigg|_{\vec{r}} & = 
-a_i \uvec{e}_{i} \times \vec{S}_{i} ,
&
\frac{\partial \av{}{H_{\star}}}{\partial \vec{e}_{i}} \bigg|_{\vec{r}} & = a_i \vec{F}_{\!i} + \frac{a_i}{e_i}	\big[(\vec{h}_{i}\times\vec{S}_{i}- \vec{C}_{i}) - (\vec{h}_{i}\times\vec{S}_{i}- \vec{C}_{i})\cdot\uvec{e}_i\,\uvec{e}_i\big],
\end{align}
with $\vec{F}_{\!i} \equiv \sum_{k} \vec{f}_{\!ik} $,
$\vec{C}_{i} \equiv \sum_{k} \vec{f}_{\!ik}\cos E_{k}  $,
$\vec{S}_{i} \equiv \sum_{k} \vec{f}_{\!ik}\sin E_{k} $ and $\vec{f}_{\!ik}$ is the gravitational force on node $k$ of wire $i$, defined by equation~\eqref{def_fik}. With these, Milankovitch's equations~\eqref{eq:Milankovitch} give equations~\eqref{EOM_r}.

\section{The relativistic precessions}
\label{sec:RelativisticPrecessions}

For the sake of completeness, in this Appendix, we spell out explicitly the $1\PN$ and $1.5\PN$ relativistic precessions induced by the central \BH\@. Following Eqs.~{(B4)} and~{(B7)} of~\cite{ToumaTremaine2009}, the respective precession velocities are given by
\begin{align}
    \dot{\vec{h}}_{\rel} &= \dot{\vec{h}}_{\rel}^{1\PN}  + \dot{\vec{h}}_{\rel}^{1.5\PN} ,
    &
    \dot{\vec{e}}_{\rel} &= \dot{\vec{e}}_{\rel}^{1\PN}  + \dot{\vec{e}}_{\rel}^{1.5\PN}  .
    \label{def_Jrel_Erel}
\end{align}
Here, the $1\PN$ contribution corresponds to the Schwarzschild precession that drives an in-plane (prograde) precession of the wire's pericentre, without any change in the wire's orbital plane. It reads
\begin{align}
    \dot{\vec{h}}_{\rel}^{1\PN} &= 0 , 
    &
    \dot{\vec{e}}_{\rel}^{1\PN} &= \frac{3 G \MBH \Omega_{\Kep} (a)}{c^{2} a (1 - e^{2})^{3/2}} \, \vec{h} \times \vec{e} ,
    \label{1PN_Prec}
\end{align}
with $\Omega_{\Kep} (a)$ the Keplerian frequency given by equation~\eqref{def_OmegaKep}. The $1.5\PN$ precession is the Lense--Thirring precession. It is sourced by the central \BH\@'s spin and leads to an orbital precession of the wire. It reads
\begin{align}
    \dot{\vec{h}}_{\rel}^{1.5\PN} &= \frac{2 (G \MBH)^{3/2} \Omega_{\rm Kep} (a)}{c^{3} a^{3/2} (1 - e^{2})^{3/2}} \, \vec{S} \times \vec{h} , &
    \dot{\vec{e}}_{\rel}^{1.5\PN} &= \frac{2 (G \MBH)^{3/2} \Omega_{\rm Kep} (a)}{c^{3} a^{3/2} (1 - e^{2})^{3/2}} \bigg[ \vec{S} - \frac{3 \vec{h} (\vec{h} \cdot \vec{S})}{1 - e^{2}} \bigg] \times \vec{e} ,
    \label{1.5PN_Prec}
\end{align}
with $|\vec{S}| \leq 1$ the normalised spin angular momentum of the central \BH\@.

\section{Spherical Harmonics}
\label{sec:SphericalHarmonics}

In this Appendix, we spell out explicitly our convention
and our implementation for the calculation of the solid spherical harmonics,
following definitions very similar to the ones of~\cite{Dehnen2014}.

The (complex) surface spherical harmonics are defined as
\begin{equation}
Y_{\ell}^{m} (\uvec{r}) = (-1)^{m} \sqrt{\frac{(\ell - m)!}{(\ell + m)!}} \, P_{\ell}^{m} (\cos (\theta)) \, \re^{\ri m \phi} , 
\label{def_Ylm}
\end{equation}
with $P_{\ell}^{m}$ the associated Legendre functions.
The (complex) solid spherical harmonics are defined as
\begin{align}
\Upsilon_{\ell}^{m} (\vec{r}) &= \big[ (\ell - m)! \, (\ell + m)! \big]^{- 1/2} \, r^{\ell} \, Y_{\ell}^{m} (\uvec{r}) ,
\nonumber
\\
\Theta_{\ell}^{m} (\vec{r}) &= \big[ (\ell - m)! \, (\ell + m)! \big]^{1/2} \, r^{- \ell - 1} \, Y_{\ell}^{m} (\uvec{r}) .
\label{def_Upsilon_Theta}
\end{align}
In order to work only with real-valued quantities,
we define the associated real-valued solid harmonics as
\begin{equation}
\label{def_Ulm}
U_{\ell}^{m} = \left\{ \begin{array}{rcrlr}
    \sqrt{2}\,\ImPart \big( \Upsilon_{\ell}^{|m|} \big) &=& \frac{1}{\sqrt{2} \, \ri} & \big[ \Upsilon_{\ell}^{-m} - \Upsilon_{\ell}^{- m *} \big] \qquad\qquad & \text{ if } m < 0 ,
\\[1ex]
    \RePart \big( \Upsilon_{\ell}^{|m|} \big) &=& \frac{1}{2} & \big[ \Upsilon_{\ell}^{m} + \Upsilon_{\ell}^{m *} \big] 
    & \text{ if } m = 0 ,
\\[1ex]
    \sqrt{2}\,\RePart \big( \Upsilon_{\ell}^{|m|} \big) &=& \frac{1}{\sqrt{2}} & \big[ \Upsilon_{\ell}^{m} + \Upsilon_{\ell}^{m *} \big] & \text{ if } m > 0 ,
\end{array}\right.
\end{equation}
and
\begin{equation}
\label{def_Tlm}
T_{\ell}^{m} = \left\{ \begin{array}{rcrlr}
    \sqrt{2}\,\ImPart \big( \Theta_{\ell}^{|m|} \big) &=& \frac{1}{\sqrt{2} \, \ri} & \big[ \Theta_{\ell}^{-m} - \Theta_{\ell}^{- m *} \big] \qquad\qquad & \text{ if } m < 0 ,
\\[1ex]
    \RePart \big( \Theta_{\ell}^{|m|} \big) &=& \frac{1}{2} & \big[ \Theta_{\ell}^{m} + \Theta_{\ell}^{m *} \big] 
    & \text{ if } m = 0 ,
\\[1ex]
    \sqrt{2}\,\RePart \big( \Theta_{\ell}^{|m|} \big) &=& \frac{1}{\sqrt{2}} & \big[ \Theta_{\ell}^{m} + \Theta_{\ell}^{m *} \big] & \text{ if } m > 0 ,
\end{array}\right.
\end{equation}
Let us note that this definition differs from the definition in Eqs.~{(58a)} and~{(58b)}
of~\cite{Dehnen2014}, where here we added a factor $\sqrt{2}$ to the expressions for $m \neq 0$.
With the present convention, the Legendre expansion is unchanged when using the real-valued harmonics,
i.e., one has for $ r_{1} < r_{2}$
\begin{align}
\frac{1}{|\vec{r}_{1} - \vec{r}_{2}|} & \, = \sum_{\ell , m} \Upsilon_{\ell}^{m} (\vec{r}_{1}) \, \Theta_{\ell}^{m *} (\vec{r}_{2}) 
\nonumber
\\
& \, = \sum_{\ell , m} U_{\ell}^{m} (\vec{r}_{1}) \, T_{\ell}^{m} (\vec{r}_{2}) .
\label{unchanged_Legendre}
\end{align}
However, compared to~\cite{Dehnen2014}, this prefactor will slightly affect some of the recurrence relations used to compute the real-valued harmonics and their gradients, as we will now detail. The important property of all the coming recurrences is that, for a given location $\vec{r}$, they allow for the computation of $U_{\ell}^{m} (\vec{r})$, $T_\ell^m(\vec{r})$ and their derivatives with complexity $O(\ellmax^{2})$.

\subsection{Computing $U_{\ell}^{m}$}
\label{sec:CompUlm}

The upper real-valued solid harmonics satisfy the initialisation condition
\begin{equation}
U_{0}^{0} = 1 .
\label{init_U00}
\end{equation}
They satisfy some boundary recurrence relations for $ m = \pm \ell$.
For $\ell = 1$, it reads
\begin{align}
U_{1}^{- 1} & \, = \frac{1}{2 \ell} \bigg[ y \, \sqrt{2} \, U_{0}^{0} + x \times 0 \bigg] ,
\nonumber
\\
U_{1}^{1} & \, = \frac{1}{2 \ell} \bigg[ x \, \sqrt{2} \, U_{0}^{0} - y \times 0 \bigg] ,
\label{bound_rec_U_1}
\end{align}
with $\vec{r} = (x , y, z)$,
while for $\ell \geq 2$, it reads
\begin{align}
U_{\ell}^{- \ell} & \, = \frac{1}{2 \ell} \bigg[ y \, U_{\ell - 1}^{\ell - 1} + x \, U_{\ell - 1}^{- (\ell - 1)} \bigg] ,
\nonumber
\\
U_{\ell}^{\ell} & \, = \frac{1}{2 \ell} \bigg[ x \, U_{\ell - 1}^{\ell - 1} - y \, U_{\ell - 1}^{- (\ell - 1)} \bigg] .
\label{bound_rec_U}
\end{align}
Finally, inside the boundary of the $(\ell , m)$-domain,
they satisfy the generic recurrence relation
\begin{align}
U_{\ell}^{m} & \, = \frac{2 \ell - 1}{\ell^{2} - m^{2}} \, z \, U_{\ell - 1}^{m} -  \frac{1}{\ell^{2} - m^{2}} \, r^{2} \, U_{\ell - 2}^{m} ,
\label{generic_rec_U}
\end{align}
which can also be applied to the cases $|m| = \ell - 1$,
if one follows the convention $U_{\ell -2}^{\pm (\ell - 1)} = 0$.

\subsection{Computing $T_{\ell}^{m}$}
\label{sec:CompTlm}

The lower real-valued solid harmonics satisfy the initialisation condition
\begin{equation}
T_{0}^{0} = \frac{1}{r} .
\label{init_T00}
\end{equation}
They satisfy some boundary recurrence relations for $ m = \pm \ell $.
For $\ell = 1$, it reads
\begin{align}
T_{1}^{-1} & \, = (2 \ell - 1) \bigg[ \frac{y}{r^{2}} \, \sqrt{2} \, T_{0}^{0} + \frac{x}{r^{2}} \times 0 \bigg] ,
\nonumber
\\
T_{1}^{1} & \, = (2 \ell - 1) \bigg[ \frac{x}{r^{2}}  \, \sqrt{2} \, T_{0}^{0} - \frac{y}{r^{2}} \times 0 \bigg] ,
\label{bound_rec_T_1}
\end{align}
while for $\ell \geq 2 $, it reads
\begin{align}
T_{\ell}^{- \ell} & \, = (2 \ell - 1) \, \bigg[ \frac{y}{r^{2}} \, T_{\ell - 1}^{\ell - 1} + \frac{x}{r^{2}} \, T_{\ell - 1}^{- (\ell - 1)} \bigg] ,
\nonumber
\\
T_{\ell}^{\ell} & \, = (2 \ell - 1) \, \bigg[ \frac{x}{r^{2}} \, T_{\ell - 1}^{\ell - 1} - \frac{y}{r^{2}} \, T_{\ell - 1}^{- (\ell - 1)} \bigg] .
\label{bound_rec_T}
\end{align}
Finally, inside the boundary of the $(\ell , m)$-domain,
they satisfy the generic recurrence relation
\begin{align}
T_{\ell}^{m} & \, = (2 \ell - 1) \, \frac{z}{r^{2}} \, T_{\ell - 1}^{m} - \big[ (\ell - 1)^{2} - m^{2} \big] \, \frac{1}{r^{2}} \, T_{\ell - 2}^{m} ,
\label{generic_rec_T}
\end{align}
which can also be applied to the cases $|m| = \ell - 1$,
if one follows the convention $T_{\ell - 2}^{\pm (\ell - 1)} = 0$.

\subsection{Computing $\nabla U_{\ell}^{m}$}
\label{sec:CompNablaUlm}

The gradients of the upper spherical harmonics, $\nabla U_{\ell}^{m}$,
can be determined using our previous computation of $U_{\ell}^{m}$.
For $m = 0$, we have
\begin{equation}
\nabla U_{\ell}^{0} =
\begin{pmatrix}
- \frac{1}{\sqrt{2}} \, U_{\ell - 1}^{1}
\\
- \frac{1}{\sqrt{2}} \, U_{\ell - 1}^{-1}
\\
U_{\ell - 1}^{0}
\end{pmatrix}
.
\label{gradU_m0}
\end{equation}
For $m = 1$, we use
\begin{equation}
\nabla U_{\ell}^{-1} =
\begin{pmatrix}
- \frac{1}{2} U_{\ell - 1}^{-2}
\\
\frac{1}{2} \big[ \sqrt{2} \, U_{\ell - 1}^{0} + U_{\ell - 1}^{2} \big]
\\
U_{\ell - 1}^{- 1}
\end{pmatrix}
\;\;\; ; \;\;\;
\nabla U_{\ell}^{1} =
\begin{pmatrix}
\frac{1}{2} \big[ \sqrt{2} \, U_{\ell - 1}^{0} - U_{\ell - 1}^{2} \big]
\\
- \frac{1}{2}U_{\ell - 1}^{- 2}
\\
U_{\ell - 1}^{1}
\end{pmatrix} ,
\label{gradU_m1}
\end{equation}
while for $m \geq 2$, this relation becomes
\begin{equation}
\nabla U_{\ell}^{-m} =
\begin{pmatrix}
\frac{1}{2} \big[ U_{\ell - 1}^{- (m-1)} - U_{\ell - 1}^{-(m+1)} \big]
\\
\frac{1}{2} \big[ U_{\ell - 1}^{m-1} + U_{\ell - 1}^{m + 1} \big]
\\
U_{\ell - 1}^{- m}
\end{pmatrix}
\;\;\; ; \;\;\;
\nabla U_{\ell}^{m} =
\begin{pmatrix}
\frac{1}{2} \big[ U_{\ell - 1}^{m - 1} - U_{\ell - 1}^{m+1} \big]
\\
\frac{1}{2} \big[ - U_{\ell-1}^{-(m-1)} - U_{\ell - 1}^{- (m + 1)} \big]
\\
U_{\ell - 1}^{m}
\end{pmatrix} .
\label{gradU_m2}
\end{equation}

\subsection{Computing $\nabla T_{\ell}^{m}$}
\label{sec:CompNablaTlm}

Similarly, the gradients of the lower spherical harmonics, $\nabla T_{\ell}^{m}$,
are computed using the predetermined values of $T_{\ell}^{m}$.
For $m = 0$, we have
\begin{equation}
\nabla T_{\ell}^{0} =
\begin{pmatrix}
- \frac{1}{\sqrt{2}} T_{\ell + 1}^{1}
\\
- \frac{1}{\sqrt{2}} T_{\ell + 1}^{-1}
\\
- T_{\ell + 1}^{0}
\end{pmatrix} .
\label{gradT_m0}
\end{equation}
For $m = 1$, we use
\begin{equation}
\nabla T_{\ell}^{-1} =
\begin{pmatrix}
- \tfrac{1}{2} T_{\ell + 1}^{-2}
\\
\frac{1}{2} \big[ \sqrt{2} \, T_{\ell + 1}^{0} + T_{\ell + 1}^{2} \big]
\\
- T_{\ell + 1}^{-1}
\end{pmatrix}
\;\;\; ; \;\;\;
\nabla T_{\ell}^{1} = 
\begin{pmatrix}
\frac{1}{2} \big[ \sqrt{2} \, T_{\ell + 1}^{0} - T_{\ell + 1}^{2} \big]
\\
- \frac{1}{2} T_{\ell + 1}^{-2}
\\
- T_{\ell + 1}^{1}
\end{pmatrix} ,
\label{gradT_m1}
\end{equation}
while for $m \geq 2$, this relation becomes
\begin{equation}
\nabla T_{\ell}^{-m} = 
\begin{pmatrix}
\frac{1}{2} \big[ T_{\ell + 1}^{- (m-1)} - T_{\ell + 1}^{- (m + 1)} \big]
\\
\frac{1}{2} \big[ T_{\ell + 1}^{m-1} + T_{\ell + 1}^{m + 1} \big]
\\
- T_{\ell + 1}^{- m}
\end{pmatrix}
\;\;\; ; \;\;\;
\nabla T_{\ell}^{m} = 
\begin{pmatrix}
\frac{1}{2} \big[ T_{\ell + 1}^{m - 1} - T_{\ell + 1}^{m+1} \big]
\\
\frac{1}{2} \big[ - T_{\ell + 1}^{- (m-1)} - T_{\ell + 1}^{- (m+1)} \big]
\\
- T_{\ell+1}^{m}
\end{pmatrix} .
\label{gradT_m2}
\end{equation}

\section{Pair Hamiltonian}
\label{sec:Pair}

In this Appendix, we present a simple analytical
two-body ``Pair'' Hamiltonian used to assess the performance
of our integration schemes on long timescales.
The Hamiltonian is
\begin{equation}
\label{eq:H_Pair}
H = - a \, \vec{h}_{1} \cdot \vec{h}_{2} - b \, \vec{e}_{1} \cdot \vec{e}_{2} ,
\end{equation}
where $a,b$ are constant coefficients
and we work in dimensionless units $\Lambda_{1} = \Lambda_{2} = 1$.
Following the Milankovitch equations~\eqref{eq:Milankovitch},
the equations of motion for the two wires read
\begin{subequations}
\label{eq:evol_Pair}
\begin{align}
\dot{\vec{h}}_{1} & = a \, \vec{h}_{1} \times \vec{h}_{2} + b \, \vec{e}_{1} \times \vec{e}_{2} ,
\qquad
\dot{\vec{e}}_{1} = b \, \vec{h}_{1} \times \vec{e}_{2} + a \, \vec{e}_{1} \times \vec{h}_{2} ,
\\
\dot{\vec{h}}_{2} & = - a \, \vec{h}_{1} \times \vec{h}_{2} - b \, \vec{e}_{1} \times \vec{e}_{2} ,
\qquad
\!\!\!\! \dot{\vec{e}}_{2} = - a \, \vec{h}_{1} \times \vec{e}_{2} - b \, \vec{e}_{1} \times \vec{h}_{2} .
\end{align}
\end{subequations}

In Figures~\ref{fig:errPairs_tau} and~\ref{fig:errPairs_time},
we considered 500 independent realisations where
(i) the coefficients $a,b$ were drawn uniformly in $[0,1]$;
(ii) the wires' initial orientations $\uvec{h}, \uvec{e}$
were drawn uniformly on the unit sphere;
(iii) the initial eccentricity, $e$, was drawn uniformly in $[0,1]$.
Because this Pair Hamiltonian does not involve
any numerical orbit average it allows for a simple test
of the integration schemes without any noise
stemming from the harmonic truncation (via $\ellmax$)
or nodes sampling (via $K$).

\section{Kozai--Lidov oscillations}
\label{sec:KozaiLidov}

In this Appendix, we detail the parameters
of the Kozai--Lidov simulations presented throughout the main text.
We work in dimensionless units,
setting in particular $G = 1$.
The mass of the central \BH\
is taken to be $\MBH = 10^{6}$.
The stellar cluster is composed of $N = 2$ stars.
The outer star is of mass $ m_{\mathrm{out}} = 10$,
semi-major axis $a_{\mathrm{out}}= 10$
and initial eccentricity $e_{\mathrm{out}} = 0.5$,
while for the inner star we take
$m_{\mathrm{in}} = 1$,
$a_{\mathrm{in}} = 1$
and $e_{\mathrm{in}} = 0.01$.
At the initial time,
we set the orientation of the outer orbit
to $ (\uvec{h}_{\mathrm{out}} , \uvec{e}_{\mathrm{out}} ) = (\uvec{e}_{z} , \uvec{e}_{y} )$.
Finally, for the inner star, we impose
$\uvec{h}_{\mathrm{in}} = \cos I_{\mathrm{in}} \, \uvec{e}_{z} - \sin I_{\mathrm{in}} \, \uvec{e}_{y}$
and $\uvec{e}_{\mathrm{in}} = \cos \omega_{\mathrm{in}} \, \uvec{e}_{x} + \sin \omega_{\mathrm{in}} \, \uvec{h}_{\mathrm{in}} \times \uvec{e}_{x}$,
with the initial inclination $I_{\mathrm{in}} = 60^{\circ}$
and pericentre phase $\omega_{\mathrm{in}} = 90^{\circ}$.
To perform the multipole integration,
we use $K = 100$ nodes,
$\ellmax = 10$ for the harmonic truncation,
and $\tau = 10$ for the integration timestep.
We did not account for any relativistic precessions.
Finally, in Figures~\ref{fig:errK}--\ref{fig:errEtot_lmax},
to estimate typical dispersions,
we performed 500 independent realisations
of the system where we drew
the initial inclinations and pericentre phase
of the inner star uniformly within $\pm 5^{\circ}$
around their fiducial values,
$I_{\mathrm{in}}$ and $\omega_{\mathrm{in}}$.

In order to check the validity of our multipole algorithm,
we compare it with direct integrations
of the associated 3-body problem,
as presented in Figure~\ref{fig:rebound}.
These direct integrations were performed
using the \texttt{IAS15} integrator~\citep{Rein+2015},
available through \texttt{REBOUND}~\citep{Rein+2012},
which automatically adjusts the integration timestep.
For the direct integrations,
the two stars' initial mean anomalies
were picked at random.
We considered a total of 100 different realisations,
and found that $e_{\mathrm{in}}$
and $\uvec{h}_{\mathrm{in}} \cdot \uvec{h}_{\mathrm{out}}$
only differed by $10^{-4}$ at most
over these realisations.
As such, in Figure~\ref{fig:rebound},
we could safely limit ourselves to only representing
the median values obtained over the available realisations.
Finally, we used the mappings from equations~\eqref{democratic_coordinates}
and~\eqref{def_bj_be_instant}
to infer the stars' orbital elements from their individual locations
and velocities.

\section{A typical stellar cluster}
\label{sec:PhysicalModel}

In this Appendix, we briefly detail the parameters of the fiducial model considered throughout the numerical applications. Mimicking SgrA*~\citep{Gillessen2017}, we take the mass of the central \BH\ to be $\MBH = 4 \times 10^{6} \Msun$. We assume that the initial eccentricity distribution of the stars follows a thermal distribution, $f_{e} \propto e$, within the range $ \emin \leq e \leq \emax$, with $ (\emin , \emax) = (0, 0.99)$. For the initial distribution in the stars' semi-major axes, we assume that that it follows a (truncated) power-law distribution of the form $f_{E} (E) \propto |E|^{p}$, with $E = \Ekep$ the Keplerian energy, and the power index $p < \tfrac{3}{2}$. With such a choice, the number of stars per unit $a$ is given by
\begin{equation}
    N (a) = (3 - \gamma) \frac{N_{0}}{a_{0}} \, \bigg( \frac{a}{a_{0}} \bigg)^{2 - \gamma} ,
\label{PDF_Na}
\end{equation}
with $\gamma = p + \tfrac{3}{2}$.
For our fiducial model, we assume that the stellar distribution follows $\gamma = \tfrac{7}{4}$, the slope of the expected equilibrium~\citep{BahcallWolf1977}. We also introduced $N_{0} = g (\gamma) N (<a_{0})$ with $g (\gamma) \!=\! 2^{-\gamma} \sqrt{\pi} \, \Gamma (1 \!+\! \gamma) / \Gamma (\gamma \!-\! \tfrac{1}{2})$. Here, $N (< a_{0})$ is the number of stars physically within a radius $a_{0}$, which should not be confused with $N_{0}$, the number of stars with a semi-major axis smaller than $a_{0}$.

For the fiducial model, we assume that the total stellar mass physically enclosed within the radius of influence $\rh = 2 \, \pc $ is $\MBH$. Assuming that all the stars have the same individual mass, $\mstar$, this implies that we make the choice $a_{0} = \rh = 2 \, \pc$ and $N (< a_{0}) = \MBH/\mstar$. It finally remains to carefully pick the value of $\mstar$. To do so, we fix ourselves a range of interest in semi-major axis, namely $\amin \leq a \leq \amax$, with $(\amin , \amax) = (5 , 100)$ in milliparsecs, to which the sampling of the particles' initial semi-major axes is limited. The fiducial simulations are performed with a total number of stars, $N = 10^{4}$. Following equation~\eqref{PDF_Na}, this imposes a self-consistent relation between $N$ and $\mstar$, namely
\begin{equation}
    \mstar = \frac{\MBH}{N} g (\gamma) \, \bigg[ \bigg( \frac{\amax}{a_{0}} \bigg)^{3 - \gamma} - \bigg( \frac{\amin}{a_{0}} \bigg)^{3 - \gamma} \bigg] .
    \label{choice_mstar}
\end{equation}
Such a choice ensures that despite the limited range in semi-major axis, the stellar self-consistent potential remains close to the full expected one. For the fiducial model with $N = 10^{4}$, we find $\mstar \simeq 8.63 \, \Msun$.

In all these simulations, the integration timestep gets dictated
by the timescale associated with the (fast) in-plane relativistic precessions of the wires with small semi-major axes and large eccentricities. As such, we fix our integration timestep to be
\begin{equation}
    \tau = 10^{-2} \, \taudyn \simeq 4.6 \, \yr
    \quad \text{with} \quad
    \tau_{\mathrm{dyn}} = \frac{1}{\Omega_{\rel} (\amin , \emax)} .
\label{choice_tau}
\end{equation}
where $\Omega_{\rel} \!=\! (3 G \MBH \Omega_{\Kep} (a))/(c^{2}a(1 \!-\!e^{2}))$ follows from equation~\eqref{1PN_Prec}. With such a choice, for the (fast) wire with $(a , e) = (\amin,\emax)$, the entire precession of the pericentre's phase requires
$\sim 600$ integration timesteps. Finally, we neglected the spin of the central \BH\@, so that the sole relativistic precession was the $1\PN$ Schwarzschild precession, as dictated by Appendix~\ref{sec:RelativisticPrecessions}.

It is informative to (crudely) compare
the performance of the multipole runs
with tentative direct $N$-body integrations.
Fixing the direct integration timestep
to $\tau_{\direct} = 10^{-2} / \Omega_{\Kep} (\amin)$,
multipole runs would use timesteps
that are $\tau/\tau_{\direct} \sim 200$ times larger
than those of the direct integration.
Integrating for a single timestep has a complexity
scaling like $O(N K \ellmax^{2})$ for the multipole method
and $O(N^{2})$ for direct integration.
For the present values $N=10^{4}$, $\ellmax = 10$ and $K = 100$,
we find therefore that the multipole runs
would typically be $\sim 200$ times less computationally intensive
than direct $N$-body integrations.

As discussed in Section~\ref{sec:Discussion}, we parallelise each simulation over four cores. We truncate the harmonic expansion at $\ellmax = 10$, and use $K = 100$ nodes for each wire. We performed 500 realisations of the fiducial system, each of them integrated up to $\tmax = 25 \, \mathrm{kyr}$. With these choices of parameters, each realisation took about $9 \, \mathrm{hr}$ of computation for the \texttt{MK4} scheme (equation~\ref{def_MK4}) on four cores. Figure~\ref{fig:SRR} presents measurements from the \texttt{MK4} integrations.

To measure the diffusion coefficients presented in Figure~\ref{fig:SRR}, we proceed as follows. First, as $a$ is conserved for the secular dynamics, we split the wires according the $a$-bins from Figure~\ref{fig:SRR}. For each $a$-bin, we split the wires in $20$ logarithmic bins in $h$ for $10^{-1} \leq h \leq 1$, according to $h(t \!=\! 0)$. To suppress the pollution stemming from the tails associated with large angular momentum changes, for each $a$-bin and $h$-bin, we compute $\Delta h \!=\! h(\tmax) - h(0)$, and use the fractional moments~\citep{BarOrAlexander2013}
\begin{equation}
    \{ (\Delta h)^{2} \} = \lim\limits_{\delta \to 0} \frac{\{ | \Delta h |^{\delta} \}^{2/\delta}}{2 \big( \Gamma [\tfrac{1 + \delta}{2}] / \!\sqrt{\pi} \big)^{2/\delta}} ,
\label{fractional_moments}
\end{equation}
with $\{ \,\cdot\, \}$ the ensemble average over realisations. In practice, we used $\delta \!=\! 10^{-3}$, and finally set $D_{hh} \!=\! \{ \Delta h)^{2} \} / \tmax$. We emphasise that we did not perform any linear fits in time and only computed finite-time diffusion coefficients. Finally, to estimate the associated errors, we proceed by bootstrap resamplings, and in Figure~\ref{fig:SRR} we represent the $16\%$ and $84\%$ error levels.

In the same Figure~\ref{fig:SRR}, we also represent the theoretical predictions for the finite-time diffusion coefficients computed following~\cite{BarOrFouvry2018} and the associated code \texttt{scrrpy}. In that prediction, we considered a maximum harmonic number given by $\ellmax = 10$. In order to match the exact setup considered here, we made two additional modifications to \texttt{scrrpy} compared to~\cite{BarOrFouvry2018}. First, we restricted the range of $a$ and $e$ of the underlying stellar cluster to $\amin \leq a \leq \amax$ and $\emin \leq e \leq \emax$. Second, rather than having an exact resonance condition on the precession frequencies, we computed finite-time diffusion coefficients, following equation~{(24)} of~\cite{BarOrFouvry2018}. This last modification is an important contributor to the agreement seen in Figure~\ref{fig:SRR}.

\end{document}